\documentclass[12pt]{iopart}

\usepackage{iopams}  
\usepackage{graphicx}

\def\fun#1#2{\lower3.6pt\vbox{\baselineskip0pt\lineskip.9pt
\ialign{$\mathsurround=0pt#1\hfil##\hfil$\crcr#2\crcr\sim\crcr}}}
\def\lap{\mathrel{\mathpalette\fun <}}
\def\gap{\mathrel{\mathpalette\fun >}}
\def\mh{M_{\bullet}}
\def\msun{M_\odot}
\def\ms{\mh\textrm{--}\sigma}
\def\rh{r_\mathrm{h}}
\def\beq{\begin{equation}}
\def\eeq{\end{equation}}
\def\sbh{SBH}
\def\sbhs{SBHs}
\def\rg{r_\mathrm{g}}
\def\SgrA{Sgr A*}

\begin{document}

\title{Loss Cone Dynamics}

\author{David Merritt}

\address{School of Physics and Astronomy, Rochester Institute of Technology, Rochester, New York 14623, USA}
\ead{merritt@astro.rit.edu}
\begin{abstract}
Supermassive black holes can capture or disrupt stars that come sufficiently close.
This article reviews the dynamical processes by which stars or stellar remnants
are placed onto loss-cone orbits and the implications for feeding rates.
The capture rate is well defined for spherical galaxies with nuclear
relaxation times that are shorter than the galaxy's age.
However, even the dense nucleus of the Milky Way may be less than one
relaxation time old, and this is certainly the case for more massive galaxies;
the capture rate in such galaxies is an initial-value problem with poorly-known
initial conditions and the rate can be much higher, or much lower, than the
rate in a collisionally relaxed nucleus.
In nonspherical (axisymmetric, triaxial) galaxies, torquing of orbits by the
mean field can dominate perturbations due to random encounters,  
leading to much higher capture rates than in the spherical geometry,
particularly in (massive) galaxies with long central relaxation times.
Relativistic precession plays a crucial role in mediating the capture of
compact remnants from regions very near to the black hole, by destroying the 
orbital correlations that would otherwise dominate the  torques.
The complex dynamics of relativistic loss cones are not yet well enough understood
for accurate estimates of compact-object (``EMRI'') capture rates to be made.
\end{abstract}

\maketitle

\section{Introduction}
A supermassive black hole (\sbh)  at the center of a galaxy acts like a sink, removing stars that come sufficiently close to it.
This removal can occur in one of two ways, depending on the mass
of the \sbh\ and on the properties of the star.
At one extreme, the ``star'' can itself be a gravitationally compact object:
a stellar-mass black hole or a neutron star.
For such objects, tidal stresses from the \sbh\ are unimportant, and removal
occurs only when the object finds itself on an orbit that takes it inside the
\sbh\ event horizon.
Ordinary stars can also be swallowed whole, but only if they manage
to resist being pulled apart by tidal stresses from the \sbh.
The tidal disruption radius $r_t$---the distance from the \sbh\ at which
a star would be disrupted---is defined as 
\numparts\label{Equation:Defrt}
\begin{eqnarray}
r_t &= \,\left(\eta^2\frac{M_\bullet}{m_\star}\right)^{1/3} R_\star \\
&\approx \, 1.1\times 10^{-5} \eta^{2/3}\left(\frac{\mh}{10^8\,\msun}\right)^{1/3}
\left(\frac{m_\star}{\msun}\right)^{-1/3} \left(\frac{R_\star}{R_\odot}\right)\,\mathrm{pc} 
\end{eqnarray}
\endnumparts
where $R_\star, m_\star$ are the radius and mass of the star
and $\mh$ is the \sbh\ mass. The quantity $\eta^2$
can be interpreted as  the ratio between the duration of periapsis
passage at $r_\mathrm{peri}\approx r_t$ and the hydrodynamic timescale of the star,
and it  can be calculated given the internal properties of the star.
For stars modelled as polytropes,
with gaseous equations of state $P =K\rho^{(n+1)/n}$,
$\eta$ is related to the polytropic index $n$ as follows \cite{Diener1995}:

\bigskip
\centerline{\begin{tabular}{cccccc}\hline
$n$: &
3&
2&
1.5&
1&
0\tabularnewline
$\eta$: &
0.844&
1.482&
1.839&
2.223&
3.074\tabularnewline
\hline
\end{tabular}}
\bigskip

\noindent
For the Sun, $n\approx 3$ and $\eta\approx 0.844$.
Comparing $r_t$ with the \sbh\ gravitational radius $\rg\equiv G\mh/c^2$
yields the ratio
\begin{equation}\label{Equation:DefineTheta}
\Theta\equiv\frac{r_t}{\rg}  
\approx 2.2\, \eta^{2/3} \left(\frac{\mh}{10^8\,\msun}\right)^{-2/3}
\left(\frac{m_\star}{\msun}\right)^{-1/3}
\frac{R_\star}{R_\odot}
\end{equation}
which implies that tidal disruption occurs outside of the \sbh's event
horizon for solar-type stars when $\mh\lap 10^8\,\msun$.
A more useful comparison is between $r_t$ and $r_c$, the periapsis radius
of the critical orbit that just continues inside the \sbh.
In the case of circular orbits around nonspinning (Schwarzschild)
holes it is well known that the innermost stable radius is $6\rg$,
changing to $1(9)\rg$ in the case of prograde (retrograde)
orbits in the equatorial plane of a maximally spinning (Kerr) \sbh.
Circular orbits are not terribly likely however since capture of stars in galactic nuclei
is more likely to be preceeded by scattering onto a highly eccentric orbit,
as discussed below.
For such orbits the critical angular momentum for capture by a nonrotating \sbh\
is $\sim 4G\mh/c$;
the periapsis of a Newtonian orbit with this value of $L$ is
$\sim 8\rg$, changing to $\sim 2(12) \rg$ for prograde
(retrograde) orbits around maximally rotating \sbhs\ \cite{Will2012}.
For orbits out of the \sbh\ symmetry plane, the critical angular momentum for
capture depends on the inclination as well, returing to approximately the same
value as for nonspinning holes as $\cos i \rightarrow 0$ \cite{Will2012}.
From the condition $r_t>8\rg$ we find that solar-type stars on eccentric orbits are disrupted (not swallowed) if $\mh\lap1.2 \times 10^7\msun$; disruption can occur for any 
$\mh \lap 10^8\msun$ if the star is on a less eccentric orbit, or for Kerr SBHs even more massive than $\sim 10^8\msun$. Red giants or AGB (asymptotic-giant-branch)
stars can also be disrupted (or at least tidally limited) by SBHs more massive than 
$10^8\msun$.
The relatively short lifetimes of these giant phases, plus the fact that ``disruption''
of a giant star may leave its structure nearly unchanged, complicates the calculation
of  tidal event rates in giant galaxies.

\begin{figure}[h!]
\hbox{
\hspace{-0.0em}
\includegraphics[width=0.75\linewidth,angle=90.]{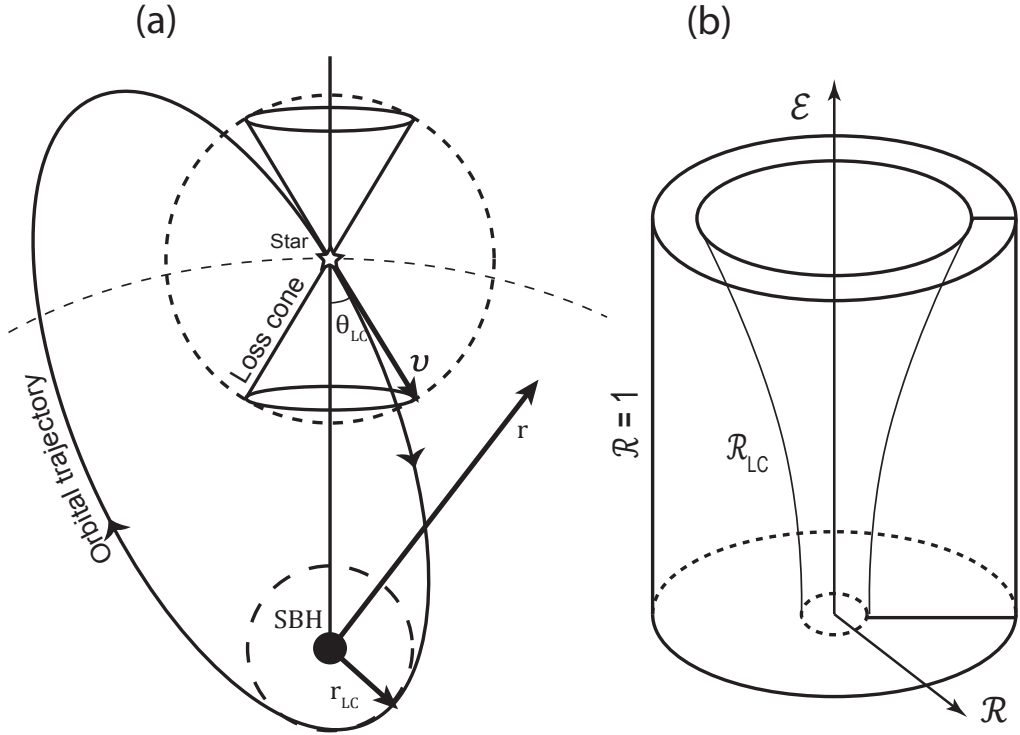} }
\caption{Two representations of the loss cone.
(a) Orbits with velocity vectors that fall within the cone
$\theta\le\theta_\mathrm{lc}$ will pass within the capture/disruption sphere
at $r= r_\mathrm{lc}$; the angle $\theta_\mathrm{lc}$ is given
approximately by equation~(\ref{Equation:Defthetalc}).
(b) In energy--angular-momentum space, the loss cone consists
of orbits with $L\le L_\mathrm{lc}$ (equation~\ref{Equation:DefLC}).
In this figure, ${\cal E}\equiv -E$ and ${\cal R}\equiv L^2/L_c(E)^2$;
${\cal R}=0$ corresponds to radial orbits and ${\cal R}=1$ to circular orbits.
The representation of the loss cone as a cylinder is motivated by
the fact that the differential equation describing \sbh\ feeding in a spherical
galaxy,
equation~(\ref{Equation:dNdtapprox}), has the same form mathematically as
the equation describing the flow of heat in an infinite cylinder.
 \label{Figure:LossCones}}
\end{figure}

The loss-cone radius $r_\mathrm{lc}$ is defined as 
the larger of the tidal disruption radius, $r_t$, or the radius of capture, $r_c$, for stars of a given type.
An orbit that just grazes the sphere at $r=r_\mathrm{lc}$ has angular momentum
\beq\label{Equation:DefLC}
L_\mathrm{lc}^2(E) = 2r_\mathrm{lc}^2\left[E-\Phi(r_\mathrm{lc})\right]
\approx 2G\mh r_\mathrm{lc}\, ;
\eeq
the latter expression assumes $|E|\ll G\mh/r_\mathrm{lc}$, that is, that the star is on an orbit with semimajor axis
much greater than $r_\mathrm{lc}$.
Orbits with $L\le L_\mathrm{lc}$
are called loss-cone orbits, and the ensemble of such
orbits is sometimes called simply the ``loss cone,''
a term that derives
from plasma physics \cite{KrallTrivelpiece1973}.
The loss cone can also be visualized as the set of
velocity vectors, at some distance $r$ from the \sbh, that are associated
with orbits that pass within $r_\mathrm{lc}$  (figure~\ref{Figure:LossCones}a).
To satisfy this condition, a star's velocity vector must lie within
a cone of half-angle $\theta_\mathrm{lc}$ that is given approximately by
\begin{eqnarray}
\theta_\mathrm{lc} &\approx&
 (r_\mathrm{lc}/r)^{1/2}, \qquad r\lap \rh, \nonumber \\
&\approx& (r_\mathrm{lc}\rh/r^2)^{1/2}, \qquad r\gap\rh
\label{Equation:Defthetalc}
\end{eqnarray}
where 
\beq\label{Equation:Definerh}
\rh\equiv \frac{G\mh}{\sigma^2} \approx 10 \left(\frac{\mh}{10^8\msun}\right)
\left(\frac{\sigma}{200\;\mathrm{km\ s}^{-1}}\right)^{-2} \mathrm{pc}
\eeq 
is the \sbh's gravitational influence
radius, defined in terms of the galaxy's central velocity dispersion $\sigma$.
(Relations \ref{Equation:Defthetalc} follow from equation \ref{Equation:DefLC},
the first after setting $v(r) \sim \sqrt{G\mh/r}$, the second after setting $v\sim\sigma$.)
Another useful definition of  ``influence radius'' is $r_\mathrm{m}$,
the radius containing a mass in stars equal to twice $\mh$:
\beq
M_\star(r<r_\mathrm{m}) = 2\mh.
\eeq
These two radii are approximately equal in galaxies with steep ($\rho\sim r^{-2}$)
nuclear density profiles; for instance, in the Milky Way, $\rh\approx r_\mathrm{m}\approx 2-3$ pc.

In a spherical galaxy, the number of stars with angular momenta
small enough to satisfy equation~(\ref{Equation:DefLC}) would ordinarily
be small;
furthermore, these stars would be removed
at the first periapsis passage, that is,
after a single orbital period.
Continued supply of stars to the \sbh\  requires some
mechanism for loss-cone repopulation: new stars
need to be transferred onto loss-cone orbits, and the rate of supply
of stars to the \sbh\ will be determined by the efficiency of the
resupply process.
An often-discussed mechanism for loss-cone repopulation is gravitational
encounters (not physical collisions) between stars, 
which cause their orbital elements to gradually evolve.
In the case of nonspherical (axisymmetric or triaxial) nuclei, feeding
rates can be high even in the absence of encounters,
since fixed torques from the distributed mass will cause orbital angular
momenta to change, on a timescale that is usually much less than the
time associated with gravitational encounters.
And even in precisely spherical nuclei, the timescale associated with
gravitational encounters, the relaxation time,
may be longer than the age of the Universe, which means that
the distribution of orbital elements need not be anywhere near to a
steady-state; in other words, loss cone repopulation
may occur at a rate that depends strongly on the ``initial conditions''.
For all of these reasons, it is not currently possible to compute capture
or disruption rates for individual galaxies with any sort of confidence.

This  review focusses on the loss-cone dynamics of idealized
systems: nuclei containing a single \sbh, and stars, or stellar remnants,
of a single mass.
Readers interested in other topics related to loss cones are referred to
a recent monograph \cite{DEGN}.

\section{Spherical symmetry}
\label{Section:LossConeSS}
\subsection{Basic variables and time scales}
\label{Section:LossConeSSBasic}
While capture or disruption occurs very near to the \sbh, it turns out that
the orbits that dominate the loss rate often extend much farther out, 
to regions where the
gravitational potential $\Phi(r)$ contains contributions from the distributed mass
as well as from the \sbh.
The appropriate orbital elements are the energy $E$
and angular momentum $L$ (both defined per unit mass); sufficiently close
to the \sbh, these can be expressed in terms of orbital semimajor axis $a$
and eccentricity $e$.
It is convenient to define the alternate variables
\beq
{\cal E}\equiv -E = -v^2/2 +\psi(r), \ \ \
{\cal R} \equiv L^2/L_c^2(E)
\eeq
where $\psi(r) = -\Phi(r)=G\mh/r - \Phi_\star(r)$ 
and $L_c({\cal E})$ is the angular momentum of
a circular orbit of energy ${\cal E}$; the scaled angular momentum ${\cal R}$
has the desirable property that it lies between 0 and 1 regardless of ${\cal E}$, 
and near the \sbh, ${\cal R}\approx 1-e^2$.
In terms of
the angle $\theta_\mathrm{lc}$ defined in figure~\ref{Figure:LossCones},
${\cal R}_\mathrm{lc} \equiv L_\mathrm{lc}^2/L_c^2\approx \theta_\mathrm{lc}^2$.
Another variable that will often be used in what follows is $\ell\equiv L/L_c=\sqrt{\cal R}\approx
\sqrt{1-e^2}$.

Once inside the loss cone, stars are lost in a time $\sim P({\cal E})$, the orbital
period at energy ${\cal E}$; in the Keplerian limit,\footnote{Non-Keplerian orbits need not be closed
and may have as many as three fundamental frequencies. Here, $P$ will always
be used to refer to the radial period, i.e. the time between periapsis passages.}
\beq
P = \frac{\pi}{\sqrt{2}} \frac{G\mh}{{\cal E}^{3/2}} =
{2\pi a^{3/2}\over \sqrt{G\mh}}
\approx 1.48 \left(\frac{\mh}{4\times 10^6\,\msun}\right)^{-1/2}
\left(\frac{a}{\mathrm{mpc}}\right)^{3/2}\,\mathrm{yr}
\label{Equation:MWperiod}
\eeq
(mpc $\equiv$ milliparsecs).
Repopulation via gravitational encounters 
occurs on a time scale that is related to the relaxation time, 
which for an infinite homogeneous medium is defined as \cite{Spitzer1987}
\begin{eqnarray}\label{Equation:DefineTr}
&&t_r = {0.34\sigma^3\over G^2 m\rho\ln\Lambda} \\
&&\approx
0.95\times 10^{10} \!\left({\sigma\over 200\,\mathrm{km\,s}^{-1}}\right)^{\!3} \!\!\left({\rho\over 10^6\,\msun\,\mathrm{pc}^{-3}}\right)^{\!-1} \!\!\left({m_\star\over \msun}\right)^{\!-1} \!\!\left({\ln\Lambda\over 15}\right)^{\!-1}\!\mathrm{yr}. \nonumber
\end{eqnarray}
Here $\rho$ is the stellar density, $\sigma$ is the one-dimensional
velocity dispersion of the stars, $m_\star$ is the mass of a single star,
and $\ln\Lambda$, the Coulomb logarithm, 
is given roughly by
\beq
\ln\Lambda\approx
\ln\left(\mh/m_\star\right) \approx \ln(N_h),
\label{Equation:DefineLogLambda}
\eeq
with $N_h\equiv \mh/m_\star$ the number of stars whose
mass equals $\mh$.
For $m_\star=\msun$ and $\mh=(0.1,1,10)\times 10^8\,\msun$,
$\ln\Lambda\approx (15,18,20)$.
\begin{figure}[h!]
\centering
\begin{tabular}{cc}
\includegraphics[width=1.00\linewidth]{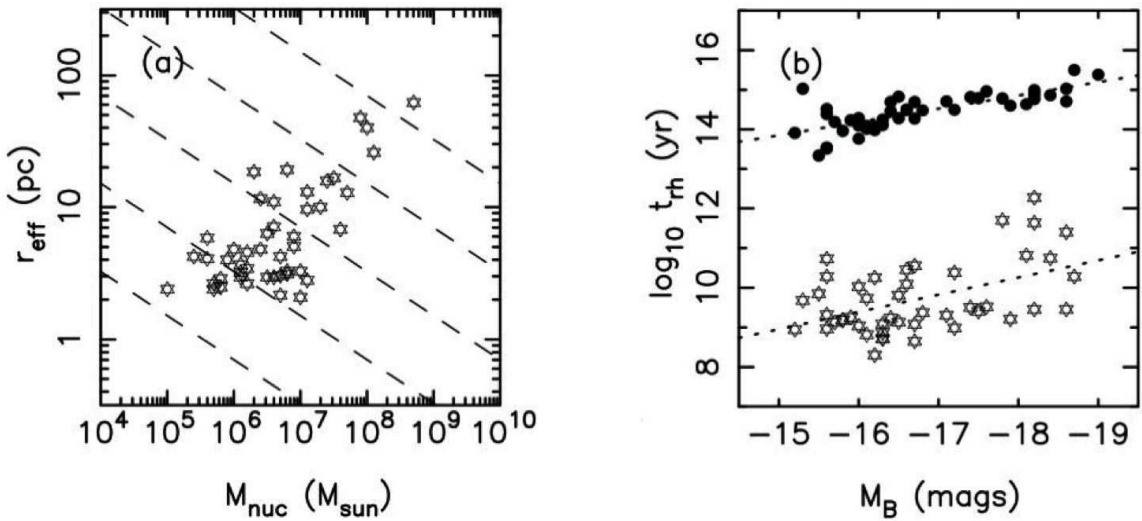}
\end{tabular}
\caption{Properties of nuclear star clusters (NSCs) in galaxies
belonging to the Virgo Galaxy Cluster \cite{Merritt2009}.
The plotted points represent all Virgo galaxies, among the 100
brightest, that have compact nuclei \cite{ACSVirgoVIII}.
{\it Left panel:} nuclear radii and masses; masses are from \cite{Seth2008a}. 
Dashed lines correspond to nuclear half-mass
relaxation times of $(10^8,10^9,10^{10}, 10^{11},10^{12})$ years
increasing up and to the right.
{\it Right panel:} half-mass relaxation times of NSCs ($\star$) and their
host galaxies ($\bullet$) plotted against absolute blue magnitude
of the galaxy.
Relaxation times were computed assuming $m_\star=\msun$.
The lower dotted line is equation (\ref{Equation:TrhNSC}).
 \label{Figure:NSCs}}
\end{figure}

In a time $\sim t_r(\rh)$, the relaxation time at the \sbh\ influence radius,
the distribution of orbital energies near the \sbh\ will have reached a statistically 
``most likely'' state due to gravitational encounters.
The corresponding density profile, the Bahcall-Wolf cusp, turns out to follow
$n(r) \propto r^{-7/4}$ at $r\lap 0.2\rh$ \cite{BahcallWolf1976}.
However it is not clear that the relaxation time in any nucleus with a known \sbh\
is short enough for a Bahcall-Wolf cusp to have formed.
The Milky Way has one of the densest nuclei known, but the relaxation time
at $\rh\approx 2.5$ pc is longer than $10^{10}$ yr \cite{Merritt2010}, 
and careful number counts of the (old) stars
inside $\rh$ reveal a flat or declining density toward the center, not a cusp
\cite{Buchholz2009,Do2009,Bartko2010}.
The galaxies most likely to contain collisionally-relaxed nuclei are those hosting
so-called nuclear star clusters (NSCs), compact sub-systems with masses 
$\sim 0.3\%$ the total galaxy mass \cite{Ferrarese2006}, and with half-mass (not central) relaxation times
that scale approximately as
\beq\label{Equation:TrhNSC}
\log_{10}\left(T_\mathrm{rh}/\mathrm{yr}\right) \approx 9.38 - 0.434\left(M_B+ 16\right)
\eeq 
with $M_B$ the absolute $B$-magnitude of the galaxy (figure~\ref{Figure:NSCs}).
These relaxation times appear to fall below 10 Gyr in galaxies with absolute magnitudes fainter than $M_B \approx -17$, or luminosities less than $\sim 4 \times 10^8 L_\odot$. 
The NSCs in these galaxies have masses below $\sim 10^7\msun$ and half-light radii 
$\lap 10$ pc. 
The NSC in the Milky Way has properties that are close to these limiting values
\cite{Schoedel2011}, 
consistent with the fact that its relaxation time is $\sim 10$ Gyr.
Unfortunately no NSC beyond the Local Group is near enough that the presence of
a Bahcall-Wolf cusp could be confirmed even if it were present; furthermore
the typical NSC may not contain a massive black hole, although
some clearly do \cite{Seth2008a}.

In stellar spheroids brighter than $\sim 10^{10} L_\odot$, which typically
do not contain NSCs, the relaxation time at the \sbh\ influence radius 
scales approximately with \sbh\ mass as \cite{MMS2007}
\begin{equation}\label{Equation:TrvsMh}
t_r(r_m) \approx 3\times 10^{11} \left(\frac{M_\bullet}{10^7\,\msun}\right)^{1.54}\,\mathrm{yr}.
\end{equation}
Relaxation times in the nuclei of brighter galaxies---or equivalently,
galaxies with \sbh\ masses greater than $\sim 10^7\,\msun$---are
probably always longer than 10\,Gyr.

\subsection{Steady-state loss rates}
\label{Section:SphericalSteadyState}

While no nucleus may be old enough for the distribution of orbital {\it energies} near
the \sbh\ to have reached a steady state, the distribution of orbital {\it angular momenta}
for values near $L_\mathrm{lc}$ may nevertheless have done so, since the associated timescale is shorter by a factor $L_\mathrm{lc}^2/L_c^2\approx {\cal R}_\mathrm{lc}^2\ll 1$ than the energy relaxation time $t_r$.
In fact it is common to follow a hybrid approach in the calculation of event rates, 
adopting whatever distribution of energies is implied by the observed $n(r)$
($f\sim {\cal E}^{\gamma-3/2}$ if $n\sim r^{-\gamma}$),
while the angular momentum distribution at each ${\cal E}$ is assumed to
have reached an approximately steady state under the influence of gravitational
encounters \cite{MagorrianTremaine1999}.

In this approach, the equation describing the evolution of the stellar phase-space
density $f({\cal R}, t; {\cal E})$ due to gravitational encounters is the one-dimensional (angular-momentum-dependent) Fokker-Planck equation \cite{DEGN}:
\begin{equation}
\label{Equation:FPRonly}
\frac{\partial N}{\partial t} =  -\frac{\partial}{\partial {\cal R}}\left(N\langle\Delta {\cal R}\rangle_t\right)
+ \frac{1}{2}\frac{\partial^2}{\partial {\cal R}^2}
\left[N\langle(\Delta {\cal R})^2\rangle_t\right].
\end{equation}
Here $N ({\cal E},{\cal R},t)d{\cal E} \,d{\cal R} = 4\pi^2 P({\cal E},{\cal R}) L_c({\cal E})^2 f({\cal E},{\cal R},t) d{\cal E} \,d{\cal R} $ is the joint distribution of orbital energies
and angular momenta, and $\langle\Delta {\cal R}\rangle$ and $\langle(\Delta {\cal R})^2\rangle$ are local, angular momentum diffusion coefficients.
The subscripts $t$ on the diffusion coefficients indicate orbit averages, for example, 
\beq
\langle\Delta {\cal R}\rangle_t \equiv \frac{2}{P}\int_{r_-}^{r_+} \frac{dr}{v_r}\langle\Delta {\cal R}\rangle 
\eeq
with $v_r$ the radial velocity along the orbit.
The local diffusion coefficients can be expressed in terms of
the velocity-space diffusion coefficient $\langle \Delta v_\perp^2\rangle$;
in the limit of small ${\cal R}$ the relations are
\beq
\label{Equation:DRsmallR}
\langle\Delta{\cal R}\rangle = \frac{r^2}{L_c({\cal E})^2}\langle\Delta v_\perp^2\rangle , \ \ \ \ 
\langle(\Delta{\cal R})^2\rangle = \frac{2r^2}{L_c({\cal E})^2}{\cal R}\langle\Delta v_\perp^2\rangle 
\eeq
and to lowest order in ${\cal R}$,
\beq\label{Equation:DRrelation}
\langle\Delta {\cal R}\rangle=\frac{1}{2}\frac{\partial}{\partial {\cal R}}
\langle\left(\Delta {\cal R}\right)^2\rangle ,
\eeq
a relation that is also valid if the local diffusion coefficients are replaced
by their orbit-averaged counterparts.
Using~(\ref{Equation:DRrelation}), equation~(\ref{Equation:FPRonly}) can be written
\beq\label{Equation:dNdtapprox}
\frac{\partial N}{\partial t} \approx
\frac{1}{2}\frac{\partial}{\partial {\cal R}}
\left[\langle\left(\Delta {\cal R}\right)^2\rangle_t \frac{\partial N}{\partial {\cal R}}\right]
 \approx {\cal D} \frac{\partial }{\partial {\cal R}}
\left({\cal R} \frac{\partial N}{\partial {\cal R}}\right),
\eeq
where
\beq
\label{Equation:DefDofE}
{\cal D}({\cal E}) \equiv \lim_{{\cal R}\rightarrow 0}
\frac{\langle(\Delta{\cal R})^2\rangle_t}{2{\cal R}}
 =\frac{2}{L_c^2({\cal E})P({\cal E})}  \int_0^{\psi^{-1}({\cal E})}\frac{r^2 dr}{v_r}
\langle(\Delta v_\perp)^2\rangle
\eeq
is an inverse, orbit-averaged relaxation time.
Equation~(\ref{Equation:dNdtapprox}), after a trivial change of variables,
 has the same mathematical form as the
equation governing transfer of heat in a cylindrical rod \cite{Ozisik1993}.
 Because of this, some authors prefer the term ``loss cylinder'' to
``loss cone,'' and that is why figure~\ref{Figure:LossCones}b adopts a cylindrical geometry.

Consider steady-state solutions.
Setting $N\propto \ln{\cal R}$ in equation~(\ref{Equation:dNdtapprox})
implies $\partial N/\partial t=0$.
As a boundary condition, it makes sense to require that $N$ fall to zero at some finite
angular momentum, say ${\cal R} = {\cal R}_\mathrm{lc}({\cal E})$.
The steady-state solution then becomes
\beq\label{Equation:NofREss}
N({\cal R}; {\cal E})=
\frac{\ln({\cal R}/{\cal R}_\mathrm{lc})}{\ln(1/{\cal R}_\mathrm{lc})+{\cal R}_\mathrm{lc}-1} {\bar N}({\cal E}),
\eeq
where
\beq\label{Equation:DefNbar}
{\bar N}({\cal E})=\int_{{\cal R}_\mathrm{lc}}^{1} N({\cal E},{\cal R}) d{\cal R}
\eeq
is a number-weighted average of $N$ over angular momentum.
${\bar N}({\cal E})$, and the corresponding phase-space density
\beq
\label{Equation:Deffbar}
{\bar f}({\cal E}) \approx \frac{{\bar N}({\cal E})}{4\pi^2 L_c^2({\cal E})P({\cal E})},
\eeq
are approximately the $N$ and $f$ that would be inferred for
an observed galaxy if it were modeled assuming an isotropic velocity distribution.

Let $F({\cal E})d{\cal E}$ be the flux of stars (number per unit time) in energy interval
$d{\cal E}$ centered on ${\cal E}$, into the loss cone.
In general,
\beq\label{Equation:DefineFluxofE}
F({\cal E}) d{\cal E} =
-\frac{d}{dt}\bigg[\int_{{\cal R}_\mathrm{lc}}^1 N({\cal E},{\cal R}) d{\cal R} \bigg] d{\cal E}.
\eeq
Substituting equation~(\ref{Equation:dNdtapprox}) into equation~(\ref{Equation:DefineFluxofE}), and requiring
$\partial N/\partial {\cal R}$ $=0$ at ${\cal R}=1$, we find
\begin{equation}
F({\cal E}) ={\cal D}({\cal E}) {\cal R}_\mathrm{lc}\left(\frac{\partial N}{\partial {\cal R}}\right)_{{\cal R}_\mathrm{lc}}
\approx {{\bar N}({\cal E}){\cal D}({\cal E})\over \ln(1/{\cal R}_\mathrm{lc})},
\label{Equation:fluxone}
\end{equation}
where the latter expression assumes $L_\mathrm{lc}\ll L_c$.
Equation~(\ref{Equation:fluxone}) states  that a fraction
$\sim 1/|\ln({\cal R}_\mathrm{lc})|$ of stars at energies ${\cal E}$ to ${\cal E} + d{\cal E}$
are scattered into the loss cone each relaxation time \cite{FrankRees1976}.

A subtlety arises here, since the change in a star's angular momentum over one orbital period, $\delta L$, can be comparable to $L_\mathrm{lc}$ \cite{LightmanShapiro1977}.
If this is the case, the separation of time scales on which equation~(\ref{Equation:FPRonly}) is based breaks down.
We can parametrize the goodness of the diffusive approximation in terms
of $q({\cal E})\approx (\delta L/L_\mathrm{lc})^2$.
Since $\delta L\approx (P/t_r)^{1/2}L_c$ and $t_r\approx {\cal D}^{-1}$, 
we can define $q$ as
\beq\label{Equation:Defineq}
q({\cal E}) \equiv  \frac{P({\cal E}){\cal D}({\cal E})}{{\cal R}_\mathrm{lc}({\cal E})}
\eeq
where $P({\cal E})\equiv P({\cal E}, {\cal R})_{{\cal R}\rightarrow 0}$.
Near the \sbh, orbital periods are short, and stars hardly penetrate beyond
the loss-cone boundary before they are consumed or destroyed.
At these energies, $q\ll 1$ and the phase-space density vanishes throughout the loss cone
except for a very small region near the boundary---this is the empty-loss-cone,
or diffusive, regime.
For small ${\cal E}$, on the other hand, $P$ is large and ${\cal R}_\mathrm{lc}$ is small;
at these energies it is possible for a star to diffuse across the loss cone
by gravitational encounters during a {\it single} orbital period.
Consider an orbit inside the loss cone.
At $r=r_\mathrm{lc}$ on such an orbit,
there are no stars moving in an outward direction; but if one were to follow
the orbit outward, stars from neighboring orbits would be scattered onto it.
This argument suggests that, for $q\gg 1$, even orbits inside the loss cone will be fully populated---this is the full-loss-cone, or pinhole, regime.

In terms of $q$, the flux in the diffusive regime (\ref{Equation:fluxone}) is
\beq\label{Equation:Fintermsofq1}
F^\mathrm{elc}({\cal E}) \approx \frac{q}{ \ln(1/{\cal R}_\mathrm{lc})}
\frac{{\bar N}{\cal R}_\mathrm{lc}}{P}, \ \ \ \ q\ll 1
\eeq
while in the pinhole regime, stars are supplied to the \sbh\ at the same rate
as if they simply followed their unperturbed orbits:
\beq\label{Equation:Fintermsofq2}
F^\mathrm{flc}({\cal E}) \approx \frac{{\bar N} {\cal R}_\mathrm{lc}^2}{P}
, \ \ \ \ q\gg 1
\eeq
independent of the rate of encounters.
\begin{figure}
\centering
\includegraphics[width=1.\linewidth]{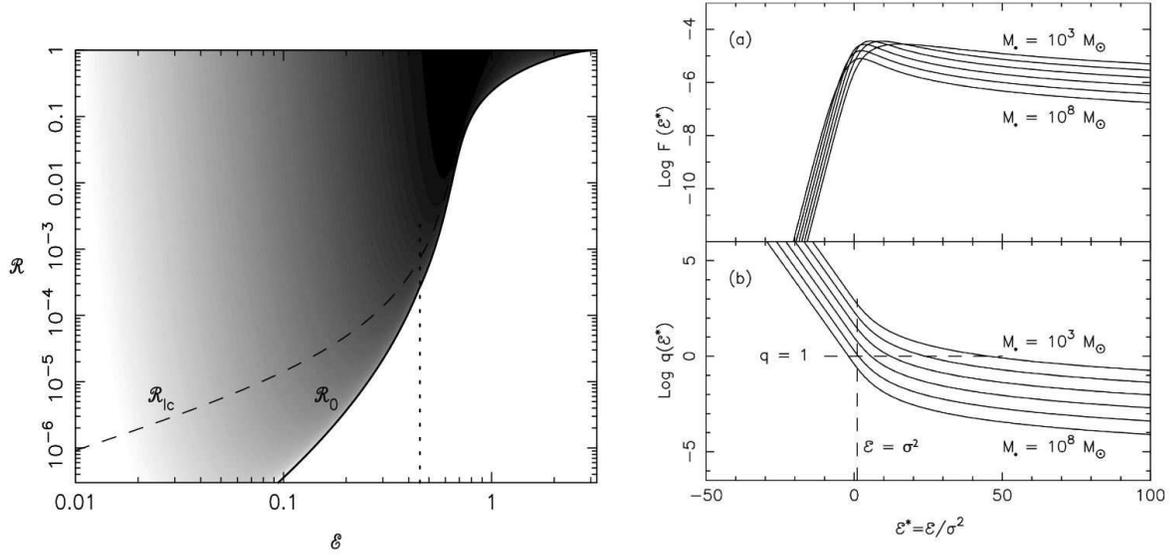}
\caption{{\it Left:} the Cohn--Kulsrud steady-state solution for $f({\cal R})$.
The gray scale is proportional to the logarithm of $f$.
${\cal R}_\mathrm{lc}$  is the dimensionless angular momentum of an orbit
that grazes the loss sphere.
The vertical dotted line marks the energy at which $q=1$;
to the left, $q\gg1$ and ${\cal R}_\mathrm{lc}\gg {\cal R}_0$
(full loss cone); to the right, $q\ll1$ and ${\cal R}_\mathrm{lc}\ll {\cal R}_0$
(empty loss cone).
{\it Right:}
(a) Energy-dependent flux of stars into the loss cone of an \sbh\ embedded in an ``isothermal'' nucleus, with density and potential given by equation~(\ref{Equation:SIS})
\cite{WangMerritt2004}.
Solar-type stars were assumed, and the $\ms$ relation was used to
relate $\sigma$ to $\mh$.
(b) The dimensionless function $q({\cal E})$
that describes the degree to which the loss cone is filled by gravitational
scattering.
For small $\mh$, most of the stars inside the \sbh\ influence sphere
are in the full-loss-cone regime.
 \label{Figure:CK}}
\end{figure}

It turns out that the flux of stars into the \sbh\ can contain substantial
contributions both from the diffusive and pinhole regimes.
This fact necessitates a quantitative understanding of how stars get into the
loss cone even at energies where the orbit-averaged approximation breaks down.
By returning to the local (radius-dependent) Fokker-Planck equation,
Cohn \& Kulsrud \cite{CohnKulsrud1978} found that the density near the
loss cone could be expressed approximately in terms of an equation like
 (\ref{Equation:NofREss}):
\beq\label{Equation:NofRECK}
N({\cal R}; {\cal E})=
\frac{\ln({\cal R}/{\cal R}_0)}{\ln(1/{\cal R}_0)+{\cal R}_0-1} {\bar N}({\cal E}),
\eeq
where ${\cal R}_0={\cal R}_0({\cal E})={\cal R}_0[q({\cal E})]$ is given by \cite{DEGN}
\beq\label{Equation:R0vsq}
{\cal R}_0(q) = {\cal R}_\mathrm{lc} e^{-q/\xi(q)}, \ \ \ \ 
\xi(q) \equiv 1 - 4\sum_{m=1}^\infty\frac{e^{-\alpha_m^2q/4}}{\alpha_m^2}
\eeq
and the $\alpha_m$ are consecutive zeros of the Bessel function $J_0(\alpha)$.
For small $q$, $\xi \approx (2/\sqrt{\pi})\sqrt{q}\approx 1.13 \sqrt{q}$; a good approximation for arbitrary
$q$ is $\xi\approx (q^2+q^4)^{1/4}$.
In the Cohn--Kulsrud solution, the phase-space density falls approximately to zero, 
not at ${\cal R}_\mathrm{lc}$, but at a smaller angular momentum ${\cal R}_0$, 
and the separation between ${\cal R}_\mathrm{lc}$ and ${\cal R}_0$ increases as
one moves farther from the \sbh, i.e. to lower ${\cal E}$.
The left panel of figure~\ref{Figure:CK} illustrates the full solution; the transition 
from empty- to full-loss-cone regimes takes place at
${\cal E}\approx {\cal E}_\mathrm{crit}$ where $q({\cal E}_\mathrm{crit}) = \left|\ln{\cal R}_\mathrm{lc}({\cal E})\right|$.
The loss-cone flux in the Cohn--Kulsrud solution (number of stars per unit time
per unit energy) can be expressed 
in terms of the full-loss-cone flux defined in equation (\ref{Equation:Fintermsofq2}) as
\beq\label{Equation:FluxAgain}
F({\cal E})\approx q({\cal E})\frac{F^\mathrm{flc}({\cal E})}{\ln(1/{\cal R}_0)}
\eeq
which, in the large-$q$ and small-$q$ limits, has the expected forms
\numparts
\begin{eqnarray}\label{Equation:Fflccases}
F/F^\mathrm{flc} &\approx&
q|\ln{\cal R}_\mathrm{lc}|^{-1},  \ \ \ \ q\ll -\ln{\cal R}_\mathrm{lc} \\
&\approx& 1, \ \ \ \ \ \ \ \ \ \ \ \ \ \ \ \ q\gg -\ln{\cal R}_\mathrm{lc}.
\end{eqnarray}
\endnumparts

Given the Cohn--Kulsrud solution for $f({\cal R})$, 
the loss rate can be computed for any 
$\{\mh,n(r), m_\star, r_\mathrm{lc}\}$ using  (\ref{Equation:Defineq}-\ref{Equation:FluxAgain}).
A concrete nuclear model, and one that describes the NSCs of the Milky Way
and some other Local Group galaxies (M32, NGC 205) fairly well, 
is the singular isothermal sphere:
\beq
\rho(r) = m_\star n(r) = {\sigma^2\over 2\pi Gr^2} ,\ \ \ \ 
\psi(r) = \frac{G\mh}{r}  -2\sigma^2\ln\left({r\over \rh}\right) .
\label{Equation:SIS}
\eeq
The name derives from the fact that---in the absence of a central point mass---the 
velocity dispersion is independent of position and equal to the parameter $\sigma$.
If we are bold enough to assume that $\sigma$ is related
to $\mh$ via the $\ms$ relation \cite{FerrareseMerritt2000}, 
then $\sigma$, $n(r)$ and $\rh=G\mh/\sigma^2$ are all determined by 
the single parameter $\mh$.
Setting $r_\mathrm{lc}=r_t$ (rather than $r_c$) is appropriate for the low-mass \sbhs\
that co-exist with NSCs; using equation (\ref{Equation:Defrt}),
the tidal disruption radius can be written
\beq\label{Equation:rtoverrhSIS}
\frac{r_t}{\rh} 
\approx 1.5\times 10^{-6} \left(\frac{\eta}{0.844}\right)^{2/3}
\left({\sigma\over 100\,\mathrm{km\,s}^{-1}}\right)^{-1.24}
\left({m_\star\over\msun}\right)^{-1/3} \left({R_\star\over R_\odot}\right) .
\eeq
The right panel of figure \ref{Figure:CK} plots $F({\cal E})$ and $q({\cal E})$ in isothermal
nuclei for various values of $\mh$, assuming $m_\star=\msun$
and $R_\star=R_\odot$.
The flux exhibits a mild maximum at ${\cal E}\approx \sigma^2$ and
falls off slowly toward large (more bound) energies: in other words,
most of the disruptions occur from orbits within the gravitational influence sphere,
regardless of the value of $\mh$.
The plot of  $q({\cal E})$ shows that, for $\mh\approx 10^8\,\msun$,
the entire influence sphere lies within the empty-loss-cone regime.
As $\mh$ is reduced, more and more of the loss cone is full.

For $\{m_\star,R_\star\} = \{\msun,R_\odot\}$, the loss rate is well approximated by
\begin{eqnarray}\label{Equation:Fapp}
\dot N_\mathrm{SIS}&\equiv&\int F(E)dE \nonumber \\
&\approx& 4.3\times 10^{-4}
\left({\sigma\over 90\,\mathrm{km\,s}^{-1}}\right)^{7/2}
\left({\mh\over 4\times 10^6\,\msun}\right)^{-1}\,\mathrm{yr}^{-1} 
\end{eqnarray}
Equation~(\ref{Equation:Fapp}), combined with the $\ms$ relation,
implies $\dot N \sim \mh^{-0.25}$: consumption rates are higher in smaller galaxies---assuming that their NSCs have properties that scale in the assumed way with $\mh$.

Tidal disruption rates as high as $10^{-4}$\,yr$^{-1}$
in nuclei with $\mh=10^6\,\msun$
imply a liberated mass comparable to $\mh$ after $10$\,Gyr.
This is not necessarily a problem, since only a fraction of
the gas removed from stars is expected to find its way into the hole
\cite{Rees1990}.
Nevertheless, the high values of $\dot N$ predicted for low-luminosity
galaxies suggest that matter tidally liberated
from stars might contribute substantially to \sbh\
growth in these galaxies.
If \sbhs\ are common in dwarf galaxies and in the bulges of late-type
spiral galaxies (both very uncertain hypotheses),
these systems would dominate the total tidal flaring rate,
due both to their large numbers and to their high individual event rates \cite{WangMerritt2004}.

\subsection{Time-dependent loss rates}
Loss-cone theory
was originally directed toward understanding the observable
consequences of massive black holes at the centers of globular clusters
\cite{FrankRees1976,LightmanShapiro1977}.
Globular clusters are many relaxation times old,
and this assumption was built into the theory,
by requiring the stellar phase-space density near the hole to have
reached an approximate steady state
under the influence of gravitational encounters.
Unfortunately, the assumption of a collisionally-relaxed distribution 
is not likely to be justified in all galactic nuclei.
Figure~\ref{Figure:NSCs} suggests that only nuclei with the smallest \sbhs---smaller than the $\sim 4\times 10^6\msun$ \sbh\ at the Galactic center---are likely to have nuclear relaxation times much shorter than $10$ Gyr.

\begin{figure}[!h]
\centering
\includegraphics[width=0.75\linewidth,angle=90.]{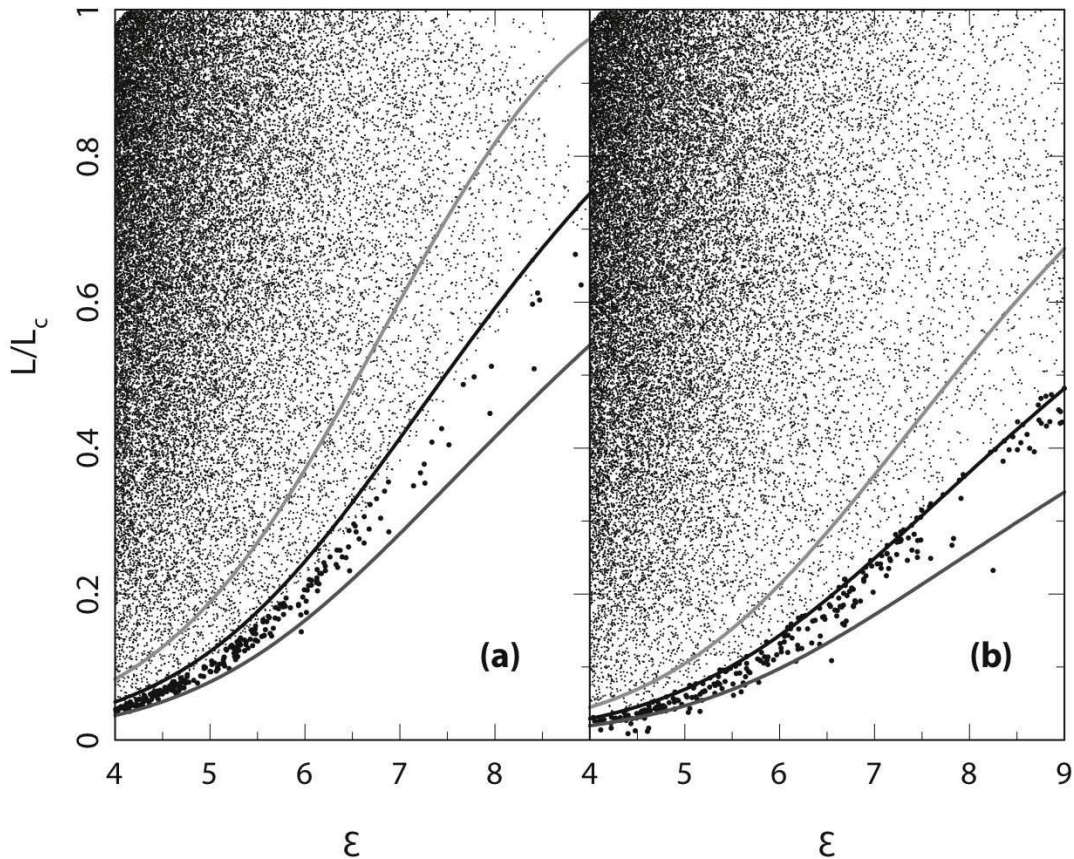}
\caption{The gap in phase space created by a binary \sbh\ at the center of a galaxy 
\cite{MerrittWang2005}. The binary mass ratios were (a) $M_2/M_1 = 1$ 
and (b) $M_2/M_1 = 1/8$. 
Curves show the angular momenta of orbits with periapses of 0.5, 1, and 2 times  $a_h$. 
The edge of the gap is approximately coincident with the middle curve in both cases, corresponding to orbits that graze the sphere $r = a_h$. The larger circles are stars that are still interacting with the binary, i.e., stars with periapses that lie within a few times $a_h$. 
These stars may still be ejected via the ``secondary slingshot'' \cite{MilosavljevicMerritt2003}.}
\label{Figure:Gap}
\end{figure}
Even in nuclei where the relaxation time is much too long for the distribution of orbital 
energies to have reached a  steady state at $r\lap \rh$, the angular momentum
distribution of stars near the \sbh\ may have evolved appreciably.\footnote{This is even 
more true at $r\ll\rh$ where resonant relaxation is effective,  as discussed below.}
The characteristic time for gravitational encounters to alter the
angular momenta of orbits with $L\lap L_0$ is
\beq\label{Equation:DeftL}
t_L\approx {L_0^2\over L_c^2} t_r.
\eeq
If $L_0$ is equal to $L_\mathrm{lc}$, the angular momentum of an orbit 
with periapsis $r_\mathrm{lc}$,  then $t_L\ll t_r$.
But giant galaxies almost universally exhibit ``cores,'' 
with sizes $10^1$--$10^2$\,pc \cite{Ferrarese1994,Lauer1995,ACSVirgoVI},
and one widely discussed model attributes cores to the ejection
of stars by a binary \sbh\ during a galaxy merger.
The massive binary creates a gap in phase space (figure \ref{Figure:Gap})
corresponding to orbits that intersected the binary at some point in its
evolution from $\Delta r \approx \rh$ to $\Delta r\approx a_h$,
where $a_h$ is the ``hard binary separation'': the separation at which the
two \sbhs\ are close enough together to eject passing stars completely
out of the nucleus.
$N$-body simulations \cite{Merritt2006} suggest $a_h\approx \nu r_m/4$, 
 where $\nu\equiv M_1M_2/(M_1+M_2)^2$ is the reduced mass
ratio of the binary and $r_m$ is the influence radius defined above.
Replacing $L_0^2$ by $2G\mh a_h$ in equation~(\ref{Equation:DeftL}),
and writing $L_c^2\approx G\mh r_m$,
appropriate for stars at a distance $\sim r_m$
from the \sbh, yields
\beq\label{Equation:tLoverTr}
{t_L\over t_r(r_m)} \approx {2a_h\over \rh} \approx \frac{\nu}{2} \approx {M_2\over 2 M_1} 
\eeq
where the last expression assumes $M_2\ll M_1$.
Combining equation (\ref{Equation:tLoverTr}) with equation (\ref{Equation:TrvsMh}) yields
\beq\label{Equation:tLoverTr2}
t_\mathrm{L} \approx 1.5\times 10^{10} \left(\frac{\nu}{0.1}\right) \left(\frac{\mh}{10^7\msun}\right)^{1.54} \mathrm{yr} .
\eeq
Equation (\ref{Equation:tLoverTr2}) suggests that the distribution of orbital
angular momenta of stars near the \sbh\
should be assumed to be gradually evolving
in nuclei with $\mh\gap 10^7\msun$.

Returning to equation~(\ref{Equation:dNdtapprox}), 
and changing variables to $\ell=\sqrt{\cal R}$,
the evolution equation can be written 
\beq
\label{Equation:HeatDiffusion}
\frac{\partial N}{\partial t} =
\frac{\mu}{\ell}\frac{\partial }{\partial \ell}
\left(\ell \frac{\partial N}{\partial \ell}\right), \ \ \ \ \mu({\cal E})\equiv \frac{{\cal D}({\cal E})}{4}.
\eeq
The assumption of diffusive evolution---i.e. that a star's angular momentum
changes very little over one orbital period---is very well satisfied
here since $a_h\gg r_t$; in other words, we are almost always in the empty-loss-cone
regime.
Equation~(\ref{Equation:HeatDiffusion}) is the heat conduction equation in cylindrical coordinates,
with radial variable $\ell$ and diffusivity $\mu$ \cite{Ozisik1993},
and the solution can be expressed in terms of a Fourier--Bessel series;
the boundary conditions are
\beq
\label{Equation:bdry}
\left. \frac{\partial N}{\partial \ell}\right|_{\ell=1}=0 \ \ \mathrm{and} \ \ 
N({\cal E},\ell)=0, \  \ell \leq \ell_\mathrm{lc}({\cal E}) = {\cal R}_\mathrm{lc}({\cal E})^{1/2}.
\eeq
\begin{figure}[h!]
\centering
\includegraphics[width=1.0\linewidth]{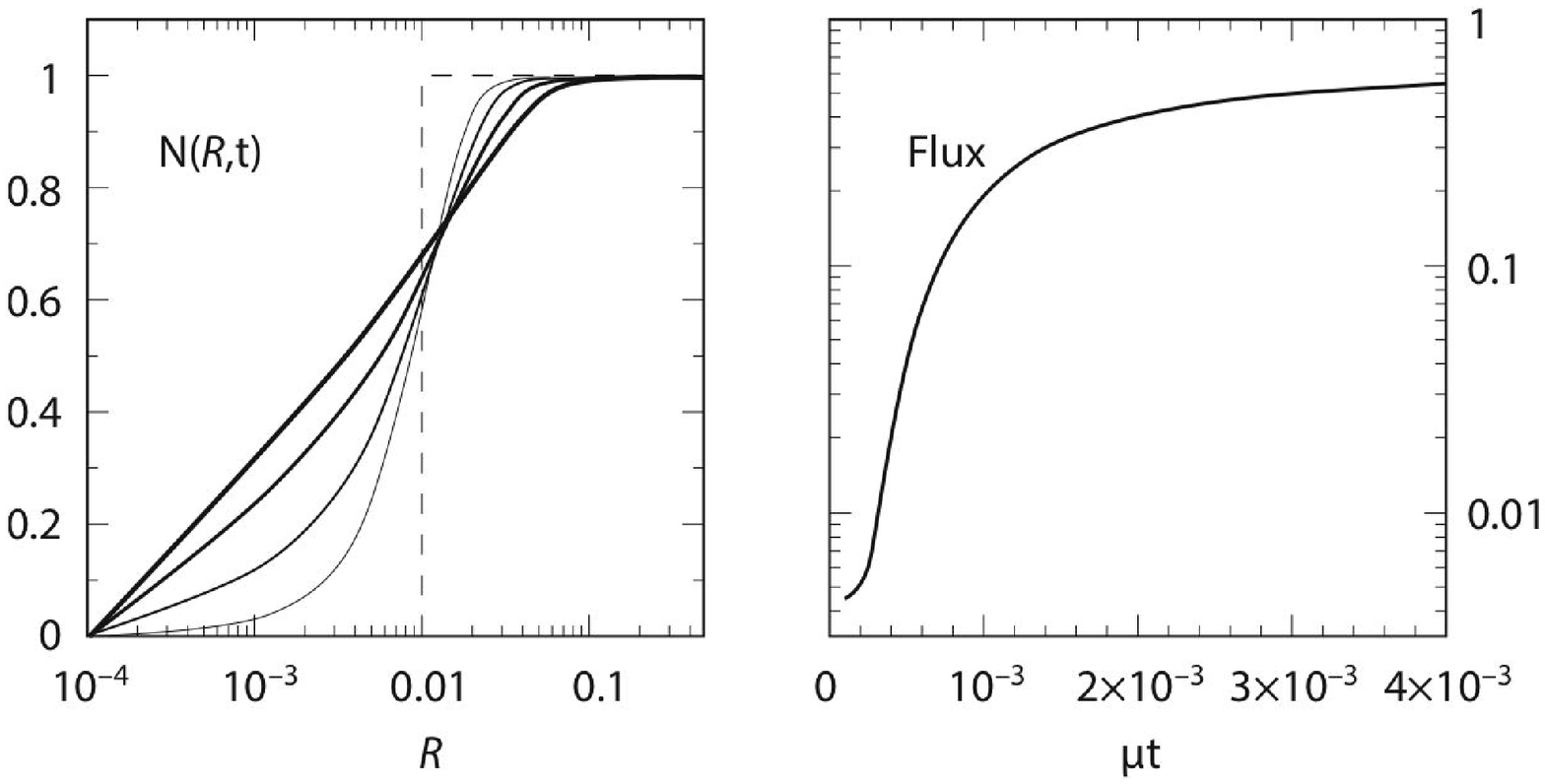}
\caption{
The left panel shows the evolution of  $N({\cal E}, {\cal R} ,t)$ at one ${\cal E}$,
computed using equation~(\ref{Equation:HeatDiffusion}).
The right panel shows the flux (per unit of ${\cal E}$) into the loss cone
at this ${\cal E}$.
The initial $N({\cal R})$ is shown at left as the dashed line:
$N({\cal E}, {\cal R},0)=0$ for ${\cal R}\le 0.01$.
The angular momentum of the loss cone was fixed at ${\cal R}_\mathrm{lc}=10^{-4}$.
In the left panel, times shown are $\mu t = (0, 0.05,0.1,0.2,0.4)\times 10^{-2}$;
line width increases with time.
The steady-state solution is nearly reached in a time of $\sim 10^{-2}t_r$,
consistent with the estimate of equation~(\ref{Equation:DeftL}).
\label{Figure:tdLC}}
\end{figure}
Figure \ref{Figure:tdLC} illustrates the evolution described by these equations.
The initially steep phase-space gradients decay
on the expected timescale of $\sim {\cal R}_0 t_r\sim 10^{-2} \mu^{-1}$.
At the final time, $N({\cal R})$ has nearly attained the exponential form
expected for the steady-state solution outside of an empty loss cone,
equation~(\ref{Equation:NofREss}).

\begin{figure}[!h]
\begin{center}
\begin{tabular}{cc}
\includegraphics[width=0.85\linewidth,angle=90.]{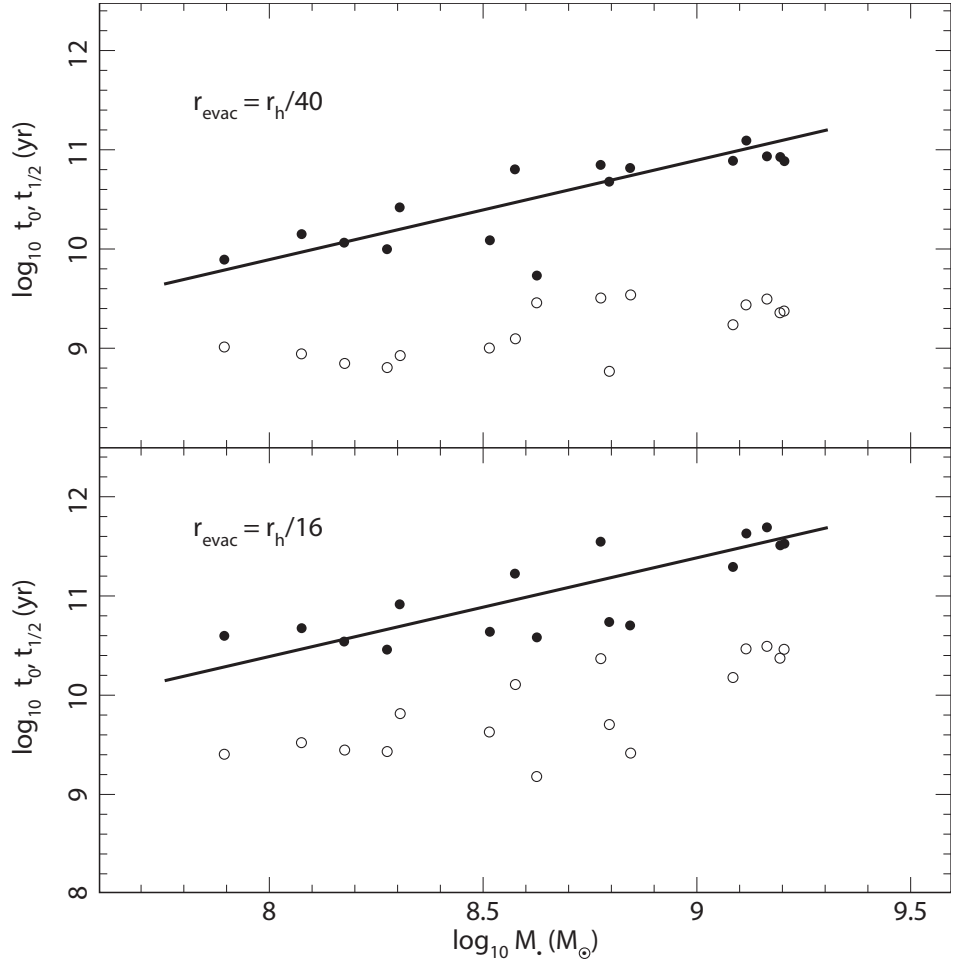}
\end{tabular}
\end{center}
\caption{
Two characteristic times associated with loss-cone
refilling in a sample of elliptical galaxies,
assuming spherical symmetry, and that initially no stars were present with periapsides
inside $r_\mathrm{evac}$ \cite{MerrittWang2005}.
$t_0$ (open circles) is the elapsed time before the first star is
scattered into the loss cone,
and $t_{1/2}$ (filled circles) is the time for the loss-cone flux to reach
$1/2$ of its steady-state value.
Solid lines are the approximate fitting function for $t_{1/2}$, equation~(\ref{Equation:fitt0}).
\label{Figure:LCrefill}}
\end{figure}
Similar calculations can be used to estimate whether loss-cone feeding
rates in observed galaxies are likely to be close to their steady-state values
\cite{MerrittWang2005}.
The initial normalization of $N({\cal R})$ at each ${\cal E}$ is fixed by the
requirement that the {\it final}, ${\cal R}$-averaged $f$ be equal to the
$f$ inferred from the galaxy's luminosity profile, assuming velocity isotropy.
Figure~\ref{Figure:LCrefill} shows the results for a sample of elliptical
galaxies, and for two values of  $r_\mathrm{evac}$, assuming that stars are initially absent
from orbits with periapsides inside $r_\mathrm{evac}$.
The values of $r_\mathrm{evac}$ are roughly what would be expected if
the current \sbh\ were preceded by a binary with mass ratio of $0.1$ or $1$.
Two characteristic times are plotted: the elapsed time before
a single star would be scattered into the loss cone; and the time
before the flux (integrated over energies) reaches one half of its steady-state value.
The latter time is found to be given roughly by
\beq
{t_{1/2}\over 10^{11}\,\mathrm{yr}} \approx
4\frac{r_\mathrm{evac}}{\rh} {M_\bullet\over 10^8\,\msun}.
\label{Equation:fitt0}
\eeq
Evidently, it would be dangerous to assume steady-state feeding rates
in galaxies with \sbhs\ more massive than $\sim 10^8\,\msun$.

\section{Nonspherical nuclei}
\label{Section:Nonspherical}

If orbits were populated at some early time without regard to the presence
of a central sink, a certain number of stars would find themselves
on loss-cone orbits.
Such stars will pass inside $r_\mathrm{lc}$ simply as a consequence of their unperturbed motion.
This process is called ``orbit draining'', and in a spherical galaxy that obeys
Jeans's theorem (i.e. in which orbits are uniformly populated with respect to
phase), 
the rate of passage of stars into the \sbh\ due to orbit draining is just
equal to the full-loss-cone rate defined above.
Orbit draining is usually ignored in the context of spherical galaxies
because the number of stars initially
on loss-cone orbits is likely to have been small,
and because these stars would have been consumed after just one orbital period.
But these arguments need to be modified in the case of nonspherical nuclei.
Torques from a flattened potential cause orbital angular momenta to change,
even in the absence of gravitational encounters.
This means that---compared with a spherical nuclus---a potentially much larger
fraction of stars can be on orbits
that will eventually bring them inside the sphere of destruction.
The timescale over which a star on such an orbit passes within $r_\mathrm{lc}$
is typically long compared with {\it radial} orbital
periods, but it may still be much shorter than the timescale for gravitational
encounters to act.

\begin{figure}[!h]
\begin{center}
\begin{tabular}{cc}
\includegraphics[width=0.90\linewidth]{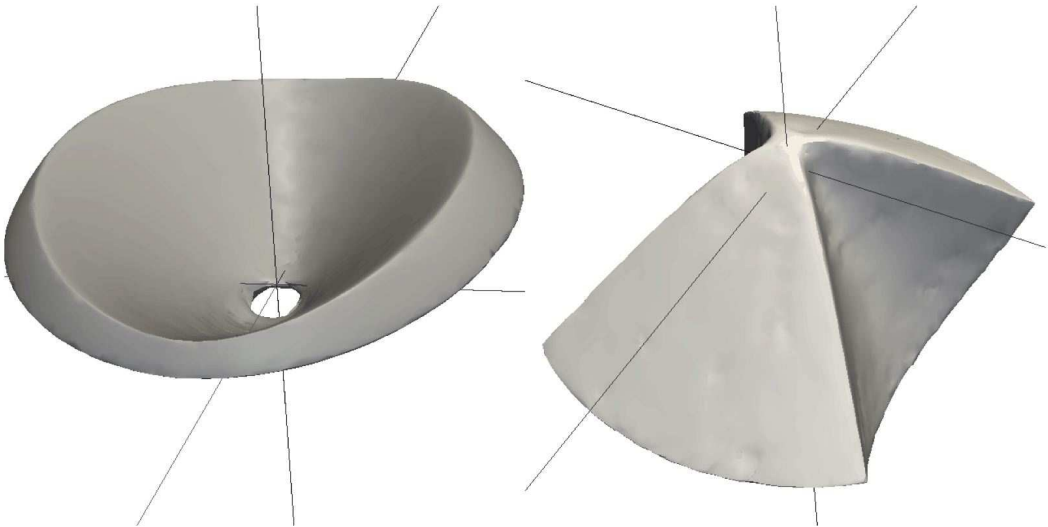}
\end{tabular}
\end{center}
\caption{Two important types of orbit that exist 
near  \sbhs\ in axisymmetric or triaxial nuclei.
{\it Left:} saucer orbit; {\it right}: pyramid orbit. 
Each figure shows the surface of the three-dimensional
volume filled by the orbit; the \sbh\ is at the origin and the short 
($z$) axis of the nucleus is indicated by the vertical line.
Saucer orbits are present in both axisymmetric and triaxial nuclei;
their excursions in $L$ are limited to $L\ge L_\mathrm{min}$, where
$L_\mathrm{min}=L_z$ in the axisymmetric case.
Pyramid orbits exist only in the triaxial geometry; they reach zero angular 
momentum at the corners of the pyramid.
Both types of orbit have counterparts obtained by reflection about a
symmetry plane of the potential.
\label{Figure:Orbits}}
\end{figure}

Nonspherical nuclei can be approximated either as axisymmetric or triaxial.
In the case of axisymmetric nuclei,
orbits conserve the energy $E$ and the component $L_z$ of the angular momentum parallel
to the symmetry axis.
In the absence of encounters, a necessary condition for a star to find its way into
the \sbh\ is $\ell_z\equiv L_z/L_c(E) <\ell_\mathrm{lc}$.
But it turns out that small-$L_z$ orbits near the \sbh\ need not conserve total 
angular momentum, even approximately; they are often ``saucers,'' 
orbits that are instantaneously close to Keplerian ellipses but whose angular 
momentum and inclination (defined with respect to the symmetry plane of the nucleus) 
oscillate in such a way that $\ell_z=\ell\cos i$ is conserved (Figure \ref{Figure:Orbits}).
\footnote{This
behavior is mathematically very similar to the behavior of orbits in the
Kozai-Lidov problem, which approximates the force from a massive object
via an axisymmetric, time-averaged potential.}
The maximum, instantaneous angular momentum of a saucer orbit turns out to be
$\ell\approx \epsilon^{1/2}\gg \ell_\mathrm{lc}$
where
\beq\label{Equation:Defineepsilon}
\epsilon\approx\frac12 (1-q)
\eeq
 and $q$ is the short-to-long axis
ratio of the stellar figure \cite{VasilievMerritt2013}.
A substantial fraction of stars with $\ell_z<\ell_\mathrm{lc}$, and with {\it instantaneous} angular momenta less than $\sim\epsilon^{1/2}$ will pass eventually within $r_\mathrm{lc}$.
If the population of low-$\ell$ orbits is not too different from the population
in an isotropic, spherical galaxy having the same radial mass distribution,
the fraction of stars at any $E$ that are
destined to pass within $r_\mathrm{lc}$ is
\beq
\sim \int_0^{\ell_\mathrm{lc}} d\ell_z \int_0^{\sqrt{\epsilon}} d\ell
\approx \sqrt{\epsilon} \ell_\mathrm{lc}
\eeq
compared with the  smaller fraction $\sim \ell_\mathrm{lc}^2$
in a spherical galaxy.
The timescale over which these orbits are drained is the longer of the
radial period and the period, $t_\mathrm{prec}$, associated with precession
through a full cycle in $\ell$ or $\cos i$; the latter time is roughly
$\sim \epsilon^{-1/2}$ times the ``mass precession time''
$t_\mathrm{M}\approx P\mh/M_\star$, i.e. the time for apsidal precession of
an orbit due to the (spherically) distributed mass.
Near the influence radius, $M_\star\approx \mh$, and in a nucleus of moderate flattening,
$\epsilon^{-1/2}t_\mathrm{M}$ will be of order or somewhat longer than $P(\rh)$.
While longer than the time required for loss-cone draining in spherical galaxies,
this time is still short enough that the saucer orbits within $\sim \rh$
would probably be drained soon after the \sbh\ is in place.

Saucer-like orbits exist also in triaxial nuclei \cite{SambhusSridhar2000}, but
much of the phase space in triaxial potentials is occupied
by an additional family of orbits: the pyramid orbits \cite{MerrittValluri1999}, 
which can be modeled as eccentric Kepler ellipses that precess, in two directions, 
about the short axis of the triaxial figure \cite{MerrittVasiliev2010}.
The angular momentum of a pyramid orbit reaches zero at the corners of the
``pyramid,'' and so {\it every} star on a pyramid orbit will eventually pass within
$r_\mathrm{lc}$\footnote{In the absence of relativistic effects.}
---unlike saucer orbits, which are limited by conservation of
$\ell_z$ to a minimum radius of periapsis.
A fraction of order $\epsilon$
of stars in a triaxial nucleus may be on pyramid orbits---larger than
the fractions $\sim \ell_\mathrm{lc}^2$ or $\sim \sqrt{\epsilon}\ell_\mathrm{lc}$ of
stars in spherical or axisymmetric nuclei that pass within $r_\mathrm{lc}$.
But the time required for a star on a pyramid orbit to reach a given
$\ell=\ell_\mathrm{lc}\ll 1$ can be much longer than for saucers,
since the angular momentum of a pyramid orbit oscillates with two independent
frequencies;
 the proportion of time that an orbit has $\ell<\ell_{lc}$ is roughly 
$(\ell_{lc}/\ell_0)^2$
where $\ell_0$ is the orbit's typical angular momentum.
The capture time is long enough that some stars on pyramid orbits, even within $\rh$,
can be expected to survive for times comparable to galaxy lifetimes.
\begin{table}[h!]
\centering
\begin{tabular}{lccc}
\hline
Geometry &Spherical & Axisymmetric & Triaxial \\
Fraction of stars\\ \phantom{x} with $\ell_\mathrm{min}<\ell_\mathrm{lc}$ & 
$\ell_\mathrm{lc}^2$ & $\sqrt{\epsilon}\ell_\mathrm{lc}$ & $\epsilon$ \\
Draining time &
$P$ & $\gap t_\mathrm{prec}$ & $\gg t_\mathrm{prec}$ \\
\hline\label{Table:Compare}
\end{tabular} 
\end{table}

Table \ref{Table:Compare} compares the different geometries.
It is  reasonable to assume that the feeding of \sbhs\ in spherical galaxies
is dominated by gravitational scattering, as discussed above,
while in triaxial galaxies gravitational encounters are of secondary
importance compared with the draining of centrophilic orbits like the pyramids.
Precisely axisymmetric nuclei are problematic.
It has been argued \cite{MagorrianTremaine1999} that feeding rates in axisymmetric galaxies
can be dominated by orbit draining, even at very late times ($\gap 10$\,Gyr) after
formation of the \sbh, implying  capture rates that are essentially the same as
in fully triaxial galaxies.
This argument assumes that there exists in axisymmetric potentials
a substantial population of chaotic orbits at $r\gg \rh$
and that these orbits are not drained at some early time.
Whether or not these assumptions are correct, 
gravitational {\it scattering} of stars into the loss cone
in axisymmetric nuclei will be affected by the presence of the saucers, even after
draining is complete,
implying modestly larger ($\sim2\times$) capture rates than in  equivalent spherical nuclei
\cite{VasilievMerritt2013}.
We discuss each case in more detail below.

\subsection{Axisymmetric nuclei}
\label{Section:NonsphericalAxisym}

Orbits near the \sbh\ in axisymmetric nuclei fall into one of two families, 
the tubes and the saucers.
Except near the separatrices dividing the two families, tube orbits behave in
a manner similar to the annular orbits in spherical potentials: the amplitude
of the total angular momentum, $L$, is nearly fixed, and there is a minimum distance
of closest approach to the \sbh\ that is related to this nearly constant $L$ by
an equation similar to (\ref{Equation:DefLC}).
Saucer orbits, on the other hand, can exhibit large angular momentum variations.
Saucer orbits exist for
\beq
\ell_z\equiv \frac{L_z}{L_c(E)} < \ell_\mathrm{sep}\approx \sqrt{\epsilon}
\eeq
with $\epsilon$ defined, as above, in terms of the nuclear shape.
When $L_z$ satisfies this condition, there exists a one-parameter set of saucer
orbits at the specified $E$ defined by the third integral $H$, or equivalently by the maximum and minimum
values of $\ell$, $\{\ell_+, \ell_-\}$, reached during a single period of oscillation
in $\ell$ and $i$.
In one class of nuclear models -- in which the density falls off as a power of
radius -- the third integral is given approximately by \cite{VasilievMerritt2013}
\begin{equation}\label{Equation:HApprox}
H(\ell,\omega) = (\ell^2-\ell_z^2) 
\left(1 - \frac{\ell^2_\mathrm{sep}}{1-\ell^2_\mathrm{sep}} 
\frac{1-\ell^2}{\ell^2} \sin^2\omega \right) 
\end{equation}
with $\omega$ the argument of periapsis of the osculating Keplerian ellipse;
equation (\ref{Equation:HApprox}) shows that $H\approx \ell^2$ for tube orbits
with $\ell\gg\ell_z$, and furthermore $H>0$ for tubes and $H<0$ for saucers.

\begin{figure}[!h]
\begin{center}
\begin{tabular}{cc}
\includegraphics[width=1.0\linewidth]{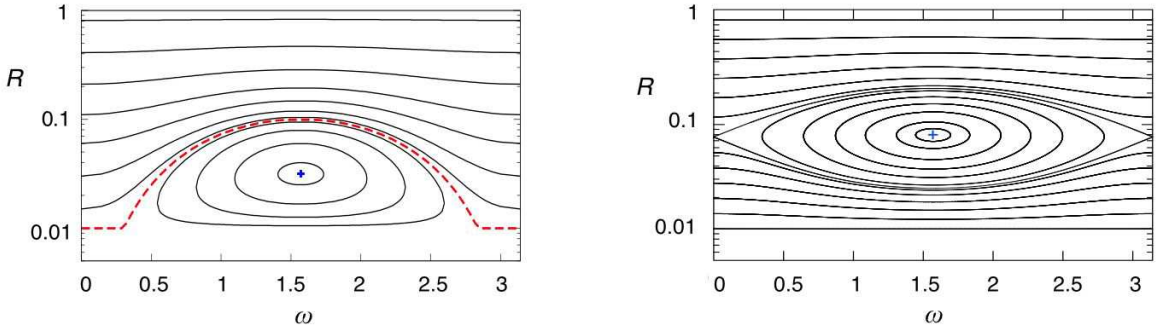}
\end{tabular}
\end{center}
\caption{Angular momentum (${\cal R}=L^2/L_c^2=\ell^2$) {\it vs}. argument of
periapsis $\omega$ for orbits near the \sbh\ in an axisymmetric nucleus. 
Each panel shows phase diagrams for orbits with the same $E$ and $L_z$ but
different values of the ``third integral'' $H$ (equation \ref{Equation:HApprox}).
{\it Left:} ${\cal R}_z = 0.01$, ${\cal R}_\mathrm{sep}=0.1$;
the fixed point orbit is marked with a cross and the separatrix with the
dashed red line. Orbits above the separatrix are tubes and orbits below 
are saucers.
{\it Right:} ${\cal R}_z = 0.01$ and $\kappa=0.02$ where $\kappa$ 
 measures the importance of relativity (eq.~\ref{Equation:Definekappa}).
An additional family of  tube orbits exists below the fixed point, 
which have sufficiently small $L$ 
that relativistic precession quenches the effects of torques due to the flattened potential. 
When $\kappa$ is increased to $\sim 1$, saucer-like orbits disappear completely.
\label{Figure:Axisym}}
\end{figure}

On the separatrix, $\ell_+=\ell_\mathrm{sep}$ and $\ell_-=\ell_z$; away from
the separatrix the variations in $\ell$ are smaller, and they are zero for the fixed-point
orbit, which has $\ell_\mathrm{f.p.}^2\approx \sqrt{\epsilon}\ell_z$,
$(\cos i)^2_\mathrm{f.p.} \approx \ell_z/\sqrt{\epsilon}$ (Figure \ref{Figure:Axisym}).
The period of oscillation in $\ell$ is very long near the separatrix but drops rapidly
away from it, with a typical value of $\sim \epsilon^{-1/2}$ times the mass precession
time $t_\mathrm{M}$:  longer than orbital periods, but probably shorter
than the timescale for gravitational encounters to change $L$ or $L_z$.

\begin{figure}[!h]
\begin{center}
\begin{tabular}{cc}
\includegraphics[width=1.0\linewidth]{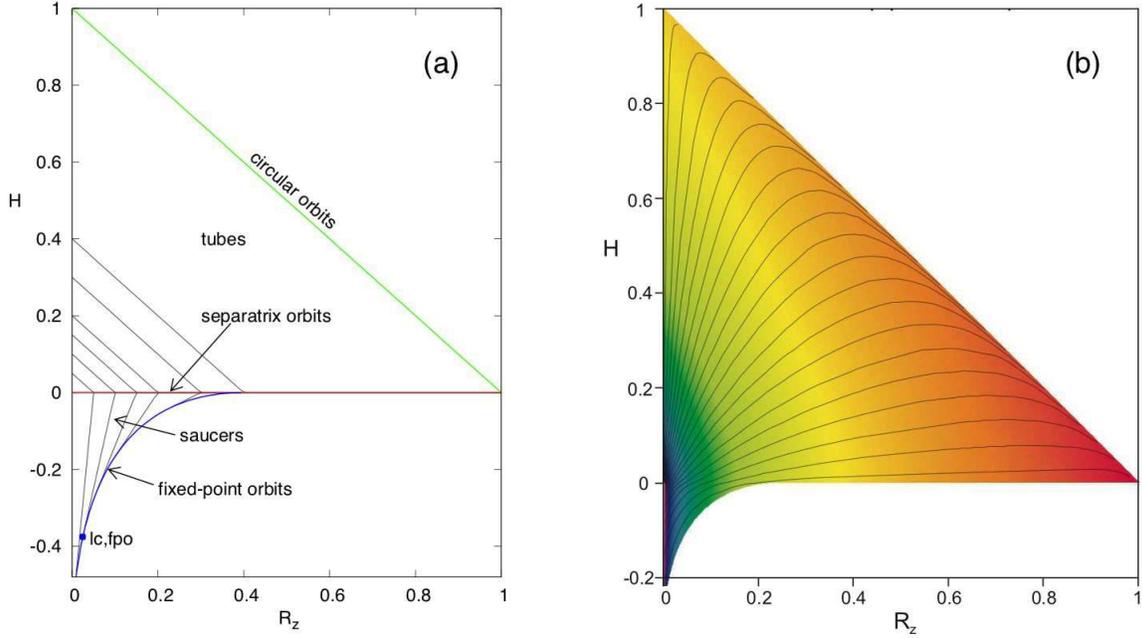}
\end{tabular}
\end{center}
\caption{(a) Phase plane in (${\cal R}_z, H$) coordinates for orbits of a
given energy in an axisymmetric nucleus;
${\cal R}_z=L_z^2/L_c^2(E) = \ell_z^2$ and $H$ is the ``third integral.''
The region $0<H<1-{\cal R}_z$ is occupied by tube orbits; $H$ is roughly
equivalent to $L$ in this region.
The bottom left corner is occupied by saucer orbits; 
the blue curve is the locus of fixed-point saucers. 
Black lines are loci of constant minimum angular momentum. 
(b) (${\cal R}_z, H$) phase plane showing stream lines
in a quasi-stationary solution to the two-dimensional diffusion problem.
More than one-half of the flux lines end up in the saucer region (at $H<0$). 
The steady-state value of $f$ is indicated by the color \cite{VasilievMerritt2013}.
\label{Figure:HRz}}
\end{figure}

In the spherical geometry, an orbit must satisfy $L<L_\mathrm{lc}$ if the star
is to go into the \sbh.
In the axisymmetric geometry, some fraction of the orbits satisfying the weaker condition $L_z<L_\mathrm{lc}$ can be captured at each $E$; this fraction is roughly the fraction
of orbits, in a spherical potential, that would have $L< L_\mathrm{sep}$. 
Since $L_\mathrm{sep}$ is typically much greater than $L_\mathrm{lc}$, 
the number of stars available for
capture can be much larger than in the spherical case.
The ``loss wedge'' \cite{MagorrianTremaine1999} 
is defined as the set of orbits that can be captured
in the absence of relaxation; that is, orbits which, at some point in their
($\ell, \omega$) precessional cycle, attain $\ell\le\ell_\mathrm{lc}$. 
The name refers to the fact that this region
is elongated in the $L$ direction (more precisely, in the direction of the third
integral $H$) much more than in $L_z$ (Figure \ref{Figure:HRz}).

To lowest post-Newtonian order, general relativity affects the motion
by inducing apsidal (also called geodetic, de Sitter, or Schwarzschild)
precession at an orbit-averaged rate
\begin{equation}\label{Equation:Sitter}
\left|\frac{d\omega}{dt}\right|_\mathrm{S} = \frac{6\pi}{P}\frac{G\mh}{c^2 a (1-e^2)} 
= \frac{3\left(G\mh\right)^{3/2}}{c^2 a^{5/2}\ell^2}
= 3\frac{\rg}{a} \frac{\nu_r}{\ell^2}
\end{equation}
where $\nu_r\equiv 2\pi/P$ is the radial (Keplerian) frequency and $\rg\equiv G\mh/c^2$.
For orbits of a given $E$, i.e. $a$,
the effects of GR precession on the motion can be characterized in terms
of the dimensionless parameter
\beq\label{Equation:Definekappa}
\kappa(a) = \frac{3 \rg}{a} \frac{\mh}{M_\star(r<a)} ,
\eeq
approximately the ratio of the GR precession frequency to the
mass precession frequency for a low-eccentricity orbit.
Saucer-like orbits are only present for $\kappa\lap 1$, i.e. if
\beq
\frac{a}{\rg} \gap 3\frac{\mh}{M_\star(a)} 
\eeq
which for the density model of equation (\ref{Equation:SIS}) implies
\beq\label{Equation:acritAxisym}
a\gap \frac{c}{\sigma} \rg \approx 10^{-2} 
\left(\frac{\mh}{10^8 \msun}\right)
\left(\frac{\sigma}{200\;\mathrm{km\ s}^{-1}}\right)^{-1} \mathrm{pc} .
\eeq
Even when saucers are present, GR precession limits their minimum angular 
momenta to 
\beq\label{Equation:ellminsaucer}
\ell_\mathrm{min} \approx \frac{2}{3\pi} \frac{\kappa}{\epsilon}.
\eeq
Requiring that $\ell_\mathrm{min} \lesssim 4\sqrt{\rg/a}$, we obtain a condition
for a star on a saucer orbit to be captured by the \sbh:
\begin{eqnarray}\label{Equation:amincapture}
a\gtrsim \left( \frac{1}{2\pi\epsilon} \frac{\sigma}{c} \right)^{2/3} r_\mathrm{m} 
\approx 0.05 \left(\frac{\epsilon}{0.01}\right)^{-2/3} 
\left(\frac{\sigma}{200\; \mathrm{km\, s}^{-1}}\right)^{2/3} r_\mathrm{m} .
\end{eqnarray}
(Of course, at smaller radii, there may still be tube-like orbits with sufficiently
small $L$.)
The radius of equation (\ref{Equation:amincapture}) is small enough
that the effects of relativity can typically be neglected in the axisymmetric
loss-cone problem, at least in the context of tidal destruction of main sequence
stars; however GR becomes crucial at the smaller radii from which compact
stellar remants would be captured, as discussed below.

The encounter-driven flux of stars into the loss wedge (per unit of energy),
${\cal F}_\mathrm{lw}$, is given by an expression like
\begin{equation}  \label{eq_capt_rate_total_flux}
{\cal F}_\mathrm{lw} = -\int_{H_\mathrm{lc,f.p.}}^{{\cal R}_\mathrm{capt}}\!\! F^{{\cal R}_z}\, dH -
  \int_0^{{\cal R}_\mathrm{capt}} (F^{H}_\mathrm{tube} - F^{H}_\mathrm{saucer})\, d{\cal R}_z\;.
\end{equation}
Here $F^{H}$ and $F^{{\cal R}_z}$ are fluxes in the $H$ and ${\cal R}_z$ directions
respectively,
$H_\mathrm{lc,f.p.}$ is the lowest possible value of $H$ 
for orbits outside the loss wedge,
and the two terms in the last integral give the contributions to the capture rate from 
the ``downward'' flux in the $H$ direction in the tube region of phase plane and 
the ``upward'' flux in the saucer region (see Figure~\ref{Figure:HRz}). 

Just as in the spherical case, however, there are two regimes, depending on
whether the radial period $P$ is short or long compared with the time required 
for an orbit to transit the loss cone.
In the axisymmetric geometry, the latter can occur in one of two ways: by
scattering, or by orderly precession.
Assume---as turns out to be correct---that the latter is more important.
Then, if the radial period is shorter than that part of the precessional cycle for which $\ell<\ell_\mathrm{lc}$, the star will be captured in a time no longer than $t_\mathrm{prec}$. 
Stars that satisfy this condition can be said to be in the ``empty-loss-wedge'' regime.
In the opposite limit, a star that achieves $\ell<\ell_\mathrm{lc}$ while  far from 
periapsis may precess out of the loss cone before capture occurs, similar 
to what happens in the full-loss-cone case of the spherical problem. 
In this ``full-loss-wedge'' regime,  precession shuffles stars 
in angular momentum quickly enough that the loss cone remains full, 
hence the capture rate is just the instantaneous number of stars inside the loss cone 
divided by their radial period --- equivalent to the full-loss-cone draining rate in the spherical geometry.
By analogy with the spherical case, the quantity 
that characterizes the two regimes is $q_\mathrm{axi}$,
where
\begin{equation}   \label{Equation:Define_qaxi}
q_\mathrm{axi} \approx \frac{P}{t_\mathrm{prec}} \frac{L_\mathrm{sep}}{L_\mathrm{lc}}
\approx \frac{P}{t_\mathrm{M}}\frac{L^2_\mathrm{sep}}{L_\mathrm{lc}} 
\approx \frac{M_\star(a)}{\mh} \frac{\epsilon}{\ell_\mathrm{lc}} .
\end{equation}
Note that we are comparing the radial period, $P$, with the (approximate) time for
the orbit to precess through the loss cone, $(\ell_\mathrm{lc}/\ell_\mathrm{sep})t_\mathrm{prec}$.
It is easy to see that $q_\mathrm{axi} \gg 1$ at the radius of influence, 
since $P \approx t_\mathrm{M}$. 
Unlike the spherical problem, the transition from 
empty- to full-loss-wedge regimes always occurs well within the radius of influence, 
and therefore the main contribution to the total capture rate comes from the 
full-loss-wedge regime. 
Moreover, for most realistic cases $q_\mathrm{axi} \gg q$ at all radii.
In other words: changes in angular momentum near the loss cone 
boundary are determined by precession and not by relaxation.

Comparing equations (\ref{Equation:Definekappa}), (\ref{Equation:ellminsaucer})
and (\ref{Equation:Define_qaxi}) we see that
\beq\label{Equation:ellminvsq}
\frac{\ell_\mathrm{min}}{\ell_\mathrm{lc}} \approx \frac{1}{\Theta\; q_\mathrm{axi}}
\eeq
with $\Theta$ defined in equation (\ref{Equation:DefineTheta}).
Roughly speaking, for orbits in the full-loss-cone ($q_\mathrm{axi}>1$) regime,
GR will not preclude a star from reaching $\ell\le\ell_\mathrm{lc}$.

The right panel of figure \ref{Figure:HRz} shows stream lines in a quasi-stationary solution
to the two-dimensional (${\cal E}, {\cal R}$) diffusion problem \cite{VasilievMerritt2013}.
More than one-half of the stream lines end up in the saucer region ($H<0$).
We can obtain an approximate expression for the capture rate if we make
a number of simplifying assumptions: (i) the distribution function depends 
only on the two classical integrals of motion, ${\cal E}$ and ${\cal R}_z$;
(ii)  the gradient of the distribution function is almost parallel to the ${\mathcal R}_z$ 
axis in the saucer region and it is in this direction that the diffusion mostly takes place; 
(iii)  precession inside the loss wedge occurs much faster than diffusion,
i.e.  $q_\mathrm{axi}\gg 1$; (iv)  the loss wedge is uniformly populated
in phase space \cite{MagorrianTremaine1999}.

Invoking assumptions (i) and (ii),
the orbit-averaged equation describing diffusion in $L_z$ is
\beq\label{Equation:dNdtLz}
\frac{\partial N}{\partial t} = 
\frac{\partial}{\partial L_z} \left(N\langle \Delta L_z\rangle_t \right) +
\frac12\frac{\partial^2}{\partial L_z^2} \left(N\langle (\Delta L_z)^2\rangle_t \right)
\eeq
with $N=N(E,L_z)$ the joint distribution of $E$ and $L_z$.
We do not know the true dependence of $f$ on the third integral, 
but assuming that $H$ is similar to the total angular momentum $L$, we can write
$N(E,L,L_z)dEdLdL_z\approx 4\pi^2 P f dE dL_z dL$ with $P$ the radial period.
Integrating this expression with respect to angular momentum
from $0$ to $L_\mathrm{sep}$ (the saucer region) then yields
\beq\label{Equation:NofELz}
N(E,L_z)\, dE\, dL_z \approx 4\pi^2 P L_\mathrm{sep} f(E,L_z) dE \, dL_z .
\eeq
Assuming an isotropic field-star distribution, the (local) diffusion coefficients,
in the limit of small $L_z$, are \cite{EinselSpurzem1999}
\numparts\label{Equation:FPLzonly}
\begin{eqnarray}\label{Equation:FPLzonlya}
\langle \Delta L_z\rangle &=& \frac{L_z}{v}\langle\Delta v_\parallel\rangle \approx 0, \\
\langle (\Delta L_z)^2\rangle &=&
\frac{L_z^2}{v^2}\langle (\Delta v_\parallel)^2\rangle
+ \frac12 \frac{(\varpi^2v^2 - L_z^2)}{v^2}
\langle (\Delta v_\perp)^2\rangle \nonumber \\\label{Equation:FPLzonlyb}
&\approx& \frac12 \varpi^2\langle(\Delta v_\perp)^2\rangle
\end{eqnarray}
\endnumparts
where $\varpi$ is the cylindrical radius.
Recall that in the spherical geometry, diffusion in $L$ at low $L$ is determined by the
quantity (\ref{Equation:DefDofE}):
\beq
D(E) \equiv \lim_{{\cal R}\rightarrow 0}
\frac{\langle(\Delta{\cal R})^2\rangle}{2{\cal R}},
\eeq
the orbit-average of which appears in the spherical diffusion 
equation~(\ref{Equation:dNdtapprox}).
Comparing equations~(\ref{Equation:DRsmallR}) and (\ref{Equation:FPLzonlyb}),
we can write for the (local) $L_z$ diffusion coefficient
\beq
\frac{\langle(\Delta L_z)^2\rangle}{L_c^2} \approx \frac{D}{2}\frac{\varpi^2}{r^2} 
\approx \frac{D}{2}\sin^2\theta
\eeq
where $\theta$ is the (instantaneous) colatitude.
We desire an orbit-averaged expression for this coefficient.
For a single, eccentric orbit, $\theta$ is nearly independent of
the mean anomaly, and so the averaging would be carried out with respect to
 $\omega$ and $i$.
Since we are seeking an estimate of the typical diffusion rate for saucer orbits
of specified $E$ and $L_z$, an additional averaging is required with respect
to the third integral.
In the absence of detailed knowledge about the distribution over that integral,
we simply
assume that $\theta$ is a uniformly populated variable over its allowed range.
Making use of the fact that $\theta$ varies nearly from $0$ to $\pi$ for
saucers with low $L_z$ (figure~\ref{Figure:Axisym}) , we can write
\beq\label{Equation:DLzOAApprox}
\langle(\Delta L_z)^2\rangle_t  \approx \frac{{\cal D}L_c^2}{4}.
\eeq
Substituting (\ref{Equation:NofELz}) and (\ref{Equation:DLzOAApprox})
into (\ref{Equation:dNdtLz}) and setting $\partial N/\partial t=0$, we find
$f(E,L_z) = a(E) + b(E)|L_z|$: a linear dependence, rather than the
logarithmic dependence characteristic of the spherical ($E,L$) loss cone.
Finally, invoking assumption (iii)---that stars are lost instantaneously after entering
the loss wedge---one finds for the flux into the \sbh\
 \cite{MagorrianTremaine1999}
\begin{equation}  \label{eq_fluxMT}
F_\mathrm{lw}(E) = 
\frac{{\mathcal{D}(E)}\,\overline N(E)} {q(E) + 4 L_c(E)/L_\mathrm{sep}(E)} 
\end{equation}
where $\overline{\mathcal N} \equiv \sqrt{2}\pi^3G^3\mh^3(-E)^{-5/2}\overline{f}$.

Comparing the fluxes in the spherical and axisymmetric geometries,
\begin{equation}\label{Equation:FcasesSA}
F \approx F^\mathrm{max}\times
\cases {
[2\ln(L_c/L_\mathrm{lc})]^{-1}, & spherical, \cr
(4L_c/L_\mathrm{sep})^{-1}, & axisymmetric\cr}
\end{equation}
where $F^\mathrm{max}(E) \equiv N(E) {\cal D}(E)$
is an esimate of the maximum rate at which stars can be driven, 
by gravitational encounters, through a constant-energy surface.
The fraction of stars of energy $E$ that are lost in one relaxation time
is $\sim 1/\ln(L_c/L_\mathrm{lc})$ in the spherical geometry and
$\sim L_\mathrm{sep}/L_c$ in the axisymmetric geometry.
The different functional dependencies reflect the fact that diffusion is
two-dimensional in the spherical case and effectively one-dimensional
in the axisymmetric case \cite{LightmanShapiro1977}.

The more complete treatment \cite{VasilievMerritt2013}
of two-dimensional ($L_z,H$) diffusion
that was the basis for figure \ref{Figure:HRz}b yields steady-state
capture rates that are better approximated by 
\begin{equation}  \label{eq_capt_rate_axi_approx}
F_\mathrm{lw}(E) = \frac{{\mathcal{D}}(E)\,\overline {N}(E)}
  {\alpha_\mathrm{axi}(E) + 2\ln(L_c/L_\mathrm{sep}) - 1 + 2\pi} 
\end{equation}
where
\begin{equation}\label{Equation:alphaaxi}
\alpha_\mathrm{axi} = 
\cases {
q/q_\mathrm{axi}, & if $q_\mathrm{axi}<1$, \cr
q, & if $q_\mathrm{axi}>1$ \cr}
\end{equation}
and $q_\mathrm{axi}$ is given by (\ref{Equation:Define_qaxi}).
For the least-bound stars, which are in the full-loss-cone regime,
$q_\mathrm{axi}\gg 1$ and $\alpha_\mathrm{axi}\approx q\gg 1$.
In this case, the capture rate does not depend on the diffusion
coefficient  ${\cal D}$ but only on the value of $L_\mathrm{lc}$,
as in the spherical problem.
In the opposite (diffusive) limit, the feeding rate is higher than in the
spherical case, but at most by a factor of a few, a prediction that is
confirmed by direct $N$-body integrations \cite{VasilievMerritt2013}.

The results obtained so far assumed regularity of the motion near the \sbh.
Near and beyond the \sbh\ influence radius,
eccentric orbits in axisymmetric potentials tend to be chaotic.
Chaotic orbits still respect the two integrals $E$ and $L_z$, 
but in principle they can fill the accessible region 
in the meridional plane, allowing them to be captured as long as
$L_z<L_\mathrm{lc}$.
Magorrian and Tremaine \cite{MagorrianTremaine1999} suggested
that feeding of \sbhs\ by chaotic orbits in axisymmetric potentials
might dominate the overall capture rate, especially in the largest galaxies
with long central relaxation times.

To a first approximation, one may assume that the chaotic orbits occupy 
a region in the $L-L_z$ plane with $L<L_\mathrm{ch}(E)$. 
The value of $L_\mathrm{ch}$ plays a role similar to $L_\mathrm{sep}$ 
inside the sphere of influence, 
and in fact is comparable to it for the same degree of flattening.
We can estimate the contribution of the chaotic orbits to the feeding rate by 
assuming also that that every chaotic orbit with a given $L_z$ can attain values 
of $L\in [L_z\ldots L_\mathrm{sep}]$ with equal probability in $L^2$.
The fraction of time such an orbit spends below the capture boundary is then
$\sim (L_\mathrm{lc}-L_z)/L_\mathrm{ch}$, and this is essentially the 
probability of being captured during one radial period.
The change with time of $f$ due to capture of stars from chaotic orbits is
given approximately by
$$
f(L_z,t;E) = f_\mathrm{init}(L_z;E)\,\exp\left[-\frac{t}{P(E)}
  \frac{L_\mathrm{lc}-L_z}{L_\mathrm{ch}(E)}\right] .
$$
The total number of chaotic orbits with $L_z<L_\mathrm{lc}$ and 
their capture rate is then given by
\numparts
\begin{eqnarray}
N_\mathrm{ch}(E,t)\,dE &=& 
 4\pi^2 \int_0^{{\cal R}capt} p(E) f({\cal R}_z,t) 
  \left(L_\mathrm{ch}/L_z-1\right) d(L_z^2/L_c^2)\, dE \nonumber\\
  &\approx& 8\pi^2 p(E) f_\mathrm{init}\, L_\mathrm{ch}L_\mathrm{lc}\,
  \frac{1-\exp(-2\tau)}{2\tau} \,dE  , \label{eq_Nchaotic} \\
F_\mathrm{ch}(E,t)\,dE &=&
  4\pi^2p(E) f_\mathrm{init} \frac{1}{P} \left(\frac{L_\mathrm{lc}}{L_c}\right)^2 \,
  \frac{1-(2\tau+1)\exp(-2\tau)}{2\tau^2} \,dE ,  \nonumber \\
\label{eq_Fchaotic}
\end{eqnarray}
\endnumparts
where $p(E)\equiv 2^{-3/2}\pi(G\mh)^3(-E)^{-5/2}$ and
\begin{equation}  \label{eq_Tdrainch}
\tau \equiv t/T_\mathrm{drain} \;,\quad 
  T_\mathrm{drain} \equiv 2P(L_\mathrm{ch}/L_\mathrm{lc}) .
\end{equation}
If we identify $f_\mathrm{init}$ with the initial value 
of the distribution function in the loss cone,
the capture rate is initially equal to the draining rate 
of a uniformly populated loss cone, equation (\ref{Equation:Fintermsofq2}).
On the other hand, the draining time 
depends on $L_\mathrm{ch}$
since the number of stars in the chaotic region 
is $2L_\mathrm{ch}/L_\mathrm{lc}$ times larger than the number of stars in the loss cone, 
therefore the draining time is longer than the radial period by the same factor.
Identifying $L_\mathrm{ch}$ with $L_\mathrm{sep}$, 
the draining time becomes
\numparts
\begin{eqnarray}
T_\mathrm{drain} &\approx& P \left(\frac{2\epsilon a}{r_\mathrm{lc}}\right)^{1/2}\\
&\approx& 6\times 10^{10} \left(\frac{\epsilon}{\Theta}\right)^{1/2}
 \left(\frac{a}{10^2\,\mathrm{pc}}\right)^2\left(\frac{\mh}{10^8\msun}\right)^{-1} \mathrm{yr},
\end{eqnarray}
\endnumparts
potentially longer than a Hubble time.
At times much longer than the draining time, the capture rate declines as $t^{-2}$.
(For regular orbits within the influence sphere, 
the draining rate declines as $t^{-3}$, 
but the draining {\it time} for saucer orbits is probably much shorter than a Hubble time.)
It follows that the capture rate for stars on chaotic orbits can remain high  
even in the absence of relaxation, provided that the initial value 
$f_\mathrm{init}$ of the distribution function inside the chaotic region was not 
much different from its value in an isotropic galaxy.
Whether this is likely to be true is open to debate; among other things, 
$f_\mathrm{init}$ must
depend on the details of the galaxy formation process (dissipative {\it vs.}
dissipationless), the prior evolution of a binary \sbh\ (which might have
emptied out a large part of the loss wedge via the gravitational slingshot), etc.

\begin{figure}[!h]
\centering
\includegraphics[width=0.395\linewidth,angle=-90.]{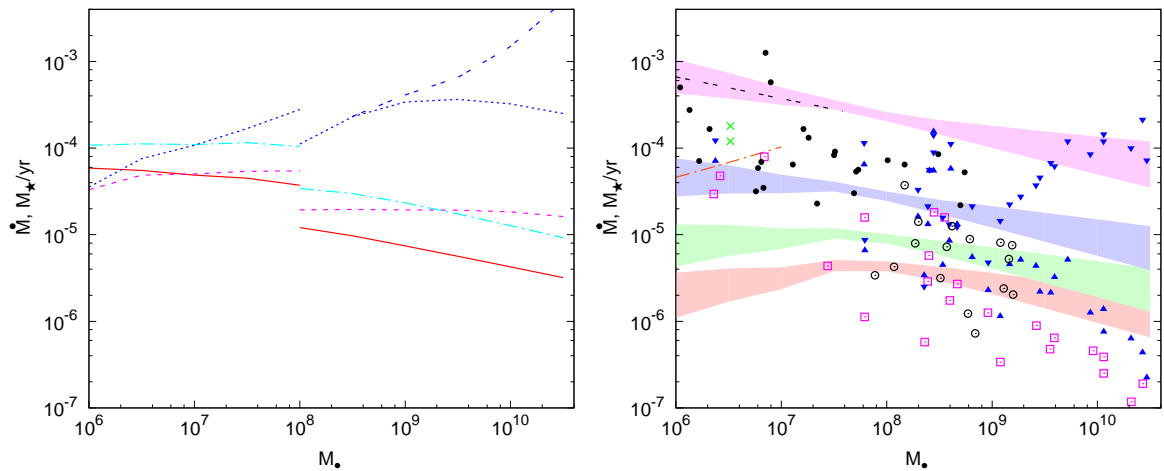}
\caption{Estimates of the capture rate $\dot M$ in spherical and axisymmetric galaxies
\cite{VasilievMerritt2013}.
Left panel shows stationary and time-dependent solutions to the Fokker-Planck equation;
details are given in the text.
Shaded bands in the right panel show the dependence of the capture rate in the spherical
geometry on uncertainties in the parameters ($\alpha,\beta$) that appear in the
$\ms$ relation; from top to bottom, the nuclear density slope is
 $\gamma\equiv -d\log\rho/d\log r=(2,1.5,1,0.5)$. 
Symbols are estimates of capture rates in individual galaxies from previous studies:
 \cite{SyerUlmer1999} purple open boxes; 
\cite{MagorrianTremaine1999} blue upward and downward triangles 
for the spherical and axisymmetric cases respectively;
 \cite{WangMerritt2004} black open and filled circles for cored and cuspy galaxies
respectively.
Black double-dashed line is equation (\ref{Equation:Fapp}), and
red dot-dashed line is based on $N$-body simulations \cite{Brockamp2011}.
\label{Figure:AxisymRates}}
\end{figure}

Figure \ref{Figure:AxisymRates} shows estimates of the capture rate in 
spherical and axisymmetric galaxies \cite{VasilievMerritt2013}.
The left panel is based on solutions to the Fokker-Planck equations 
(\ref{Equation:FPRonly}), (\ref{Equation:dNdtLz}),
assuming two sets of galaxy models:
 having either steep ($\rho\propto r^{-3/2}$) central density cusps  ($\mh\le 10^8\,M_\odot$)
or shallower ($\rho\propto r^{-1}$) ``cores'' ($\mh\ge 10^8\,M_\odot$).
The $\ms$ relation:
\begin{equation}  \label{eq_MbhSigma}
\log \mh = \alpha + \beta \log(\sigma/200\,\mathrm{km\ s}^{-1}) 
\end{equation}
with $(\alpha, \beta)=(8, 4.5)$ 
was used to relate $\mh$ to the properties of the galaxy.
Solid red and dashed purple lines are stationary and time-dependent 
rates in the spherical geometry; the time-dependent solutions assumed
initial condition with strong gradients in the distribution function near the loss 
region (as in figure \ref{Figure:tdLC}), 
which provide an upper limit to the likely rates in real spherical galaxies.
The dot-dashed and dotted blue lines show the same for axisymmetric 
galaxies with ${\cal R}_\mathrm{sep}=0.1$.
The double-dashed blue line includes an estimate of the contribution from 
draining of chaotic orbits.
Over the entire range of $\mh$, the steady-state capture rates differ by only a factor of 
$2-3$ between spherical and axisymmetric geometries, consistent with the discussion 
above. 
This result holds for any, reasonable galaxy model and depends only weakly on $L_\mathrm{sep}$.
In the time-dependent solutions, 
the approach to a steady state is slow and 
the flux at early stages is much higher than in equilibrium,
particularly in the most massive galaxies with long central relaxation times.
Also for the most massive \sbhs\ ($\mh \gtrsim 10^9\,M_\odot$) the draining time of the 
loss region  becomes comparable to the Hubble time and the capture rate is dominated by 
chaotic orbits,
reaching values up to $10^{-3}\,M_\odot$ yr$^{-1}$ 
for the largest $\mh$.
As discussed above, the time-dependent results should be considered contingent on the poorly-known
initial conditions.
 
The right panel of Figure~\ref{Figure:AxisymRates} plots variation in the 
stationary, spherical capture rate due to uncertainties in the parameters 
($\alpha, \beta$) in the $\mh-\sigma$ relation;
values for the axisymmetric geometry scale roughly in proportion. 
The shaded regions, from top to bottom, have $\gamma\equiv -d\log\rho/d\log r=(2,1.5,1,0.5)$. 
Estimates of the capture rates in individual galaxies from several previous 
studies are also plotted as symbols. 
It is clear that the scatter in the derived values is fairly large, about two orders of magnitude, 
although a general trend of decreasing  rate with increasing $\mh$ is clear.
Overall, predicted capture rates lie in the range 
$10^{-5}-10^{-4}\,M_\odot$ yr$^{-1}$ for less massive galaxies, and 
a few$\times 10^{-6}-10^{-5}\,M_\odot$ yr$^{-1}$ for galaxies with $\mh>10^8\,M_\odot$. 
(Axisymmetric) nuclear flattening may increase these numbers by a factor of few. 

One potentially important feature of axisymmetric (and triaxial) galaxies is that most stars 
are consumed in the full-loss-cone regime. This means that stars approach the 
\sbh\ with a wide distribution in periapsis radii, as opposed to ``barely touching'' the 
disruption sphere in the empty-loss-cone regime. 
One consequence is that stars can be strongly tidally distorted before disruption.
In the exchange model discussed below for the formation of the Galactic center S-stars
from binary stars, the radial distribution of the captured stars depends differently
on the initial distribution of binary separations in the empty- and full-loss-cone 
cases \cite{PeretsGualandris2010}.

\subsection{Triaxial nuclei}
\label{Section:NonsphericalTriax}

The tube and saucer orbits that characterize motion near an \sbh\ in axisymmetric
nuclei are still present in nonaxisymmetric, or triaxial, nuclei.
In fact, two families of tube orbits exist,
circulating about both the short and long axes of the triaxial figure,
as well as saucers that circulate about the short axis \cite{SambhusSridhar2000}.
Like orbits in the axisymmetric geometry, tube (saucer) orbits in triaxial potentials
respect an integral that is similar to $L$ ($L_z$) and they avoid the very center.
But triaxial potentials can also support orbits that are qualitatively different from
both tubes and saucers: ``centrophilic'' orbits that pass arbitrarily close to the \sbh\
 (figure \ref{Figure:Orbits}).

Centrophilic orbits exist even in axisymmetric nuclei, but they are restricted to
a meridional plane, that is, to a plane that contains the $z$- (symmetry) axis.
 Orbits in the meridional plane have $L_z=0$, and so conservation of $L_z$ does not impose
any additional restriction on the motion.
Perturbing such an orbit {\it out} of the meridional plane implies
a nonzero $L_z$: the orbit is converted into a saucer or a tube and again avoids the center.
But in the triaxial geometry, $L_z$ is not conserved, and it turns out that a substantial
fraction of such ``perturbed'' planar orbits will maintain their centrophilic character,
becoming pyramid orbits \cite{MerrittValluri1999}.

\begin{figure}[h!]
\centering
\includegraphics[width=0.625\textwidth,angle=-90.]{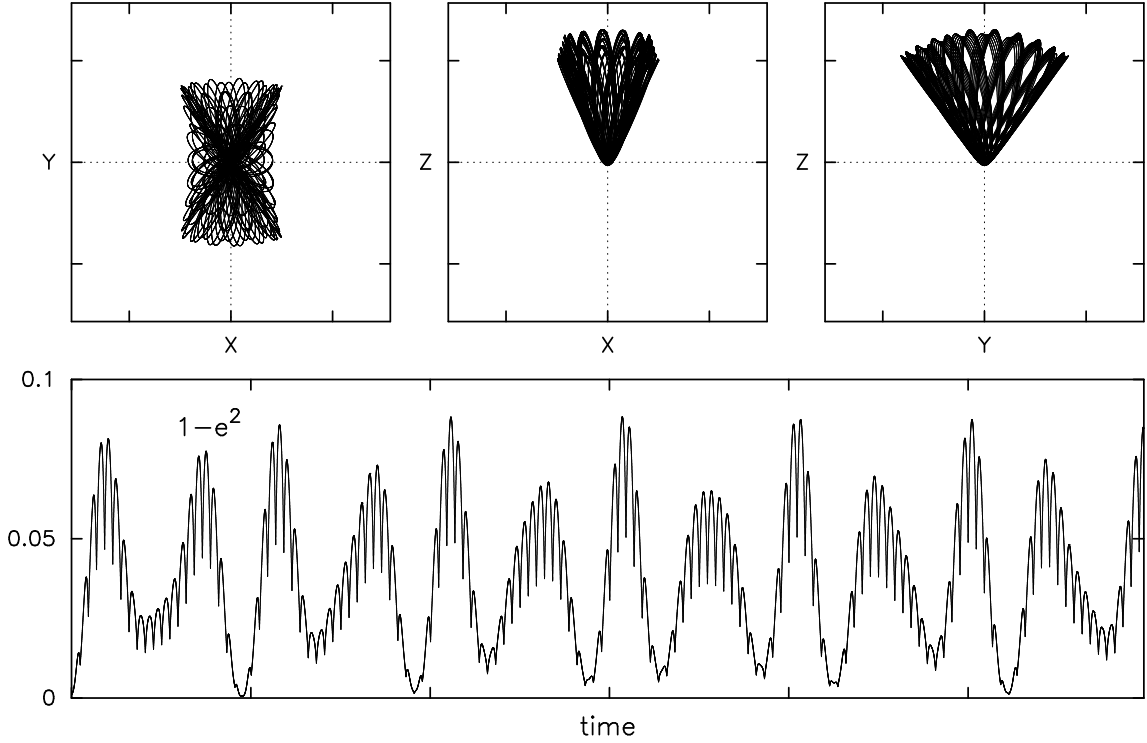}
\caption[]{
{\it Top:} A pyramid orbit, seen in three projections.
The $z$-axis is the short axis of the triaxial figure and the \sbh\
is at the origin.
{\it Bottom:} $\ell^2=1-e^2$ versus time, where $e$ is the eccentricity.
The eccentricity tends to unity when the orbit reaches the corners
of the pyramid's base. Because the frequencies of libration in $x$ and $y$ are generally
incommensurate, the corners are only reached after many libration periods. 
\label{Figure:Pyramid}
}
\end{figure}

Pyramid orbits resemble eccentric Keplerian ellipses that librate
in two directions about the short axis of the triaxial figure \cite{MerrittVasiliev2010}.
In the absence of any nonspherical component to the potential, they would
precess at a constant rate 
\beq
\label{Equation:DefineMassPrecession}
\frac{d\omega}{dt} \approx -\nu_r \sqrt{1-e^2}\left[\frac{M_\star(r<a)}{\mh}\right],
\eeq
\cite{DEGN}, 
the rate of apsidal precession due to the spherically-distributed mass $M_\star$.
But because of the torques due to the triaxial geometry, the orbit's angular momentum
varies as it precesses, from a maximum when the orbit is near the short ($z$) axis.
It is clear that -- by making the eccentricity along this axis sufficiently large --
it will always be possible to reach $e=1$ at a finite angle, since the precession
period (and hence the accumulated effect of the torques) can be made arbitrarily large.
The angles where $e=1$ correspond to the four ``corners'' of the pyramid
(figure~\ref{Figure:Pyramid}).
At a given energy $E$ there is a two-parameter family of pyramids
defined by the two non-classical integrals $U$ and $W$, or equivalently by
the ($x,y$) dimensions of the base of the pyramid.
Only orbits with peak  angular momenta (near the $z$ axis) $\ell\lap\sqrt{\epsilon}$
will be pyramids; orbits with larger $\ell$ never reach $\ell=0$ and so continue
to precess in the same direction, producing saucers or tubes.

In the large-eccentricity limit, the additional integrals can be expressed 
simply in terms of the components of a unit vector, $\boldsymbol{u}$, 
oriented toward the corners of the pyramid:
\begin{equation}
U = \nu_x^2 u_{x0}^2 + \nu_y^2 u_{y0}^2, \ \ \ \ 
W = \nu_x^4 u_{x0}^2 + \nu_y^4 u_{y0}^2
\end{equation}
where 
\begin{equation}  \label{Equation:nu0}
\nu_x = \sqrt{\frac{5\epsilon_c}{3}}\;  \;,\;\; \nu_y = \sqrt{\frac{5}{3}\left(\epsilon_c-\epsilon_b\right)}
\end{equation}
and ($\epsilon_b, \epsilon_c$) specify the axis ratios of the triaxial figure through
\beq\label{Equation:Defineepsilonbc}
\epsilon_{b,c} \simeq \left(T_{y,z} - T_x\right) \frac{\rho_t}{\rho_0}
\left(\frac{a}{r_0}\right)^{\gamma}.
\eeq
In deriving these expressions, the stellar potential was assumed to consist of
two parts: a spherical component with density $\rho_s(r) = \rho_0(r/r_0)^{-\gamma}$,
and a homogeneous triaxial bar with density $\rho_t$ and potential
\beq
 \Phi_t(x,y,z) =  2\pi\, G\rho_t\,\left(T_x x^2 + T_y y^2 + T_z z^2\right)
\label{Equation:TriaxPotential}
\eeq
where the $T_i$ are expressible in terms of the two axis ratios \cite{ChandraEllipse}.
Note that in this model, the relative contribution of the triaxial distortion to the total
stellar potential is an increasing function of radius.

Pyramids that are sufficiently eccentric (i.e. that have sufficiently compact bases)
librate in $x$ and $y$ as harmonic and uncoupled oscillators:
\begin{equation}
\label{Equation:PyramidSHO}
u_x(\tau) = u_{x0}\cos(\nu_x\tau + \phi_x), \ \ \ \ 
u_y(\tau) = u_{y0}\cos(\nu_y\tau + \phi_y)
\end{equation}
and their angular momentum varies as
\begin{equation}
\ell^2(\tau) =\ell_{x0}^2\sin^2(\nu_x\tau+\phi_x) +
\ell_{y0}^2\sin^2(\nu_y\tau+\phi_y)
\end{equation}
\label{Equation:Highe}
where $\tau\equiv \nu_\mathrm{M}t/3$ and
\numparts
\begin{eqnarray}
\ell_{x0}&=&\nu_x e_{x0}/3, \ \ \ \ \ell_{y0}=\nu_y e_{y0}/3.
\end{eqnarray}
\endnumparts
As long as the frequencies of oscillation in $x$ and $y$ are incommensurable,
the vector $(u_x,u_y)$ densely fills the  available area,
and the star comes close to the \sbh\ whenever the two
variables $(u_x,u_y)$ are simultaneously close to 1---that is, near the
corners of the pyramid.
To the extent that the simple harmonic oscillator approximation is valid,
the time-averaged probability of a given periapsis passage having $r_\mathrm{peri}<X$ is roughly
proportional to $X$.

\begin{figure}[h!]
\begin{center}
\includegraphics[width=1.0\textwidth]{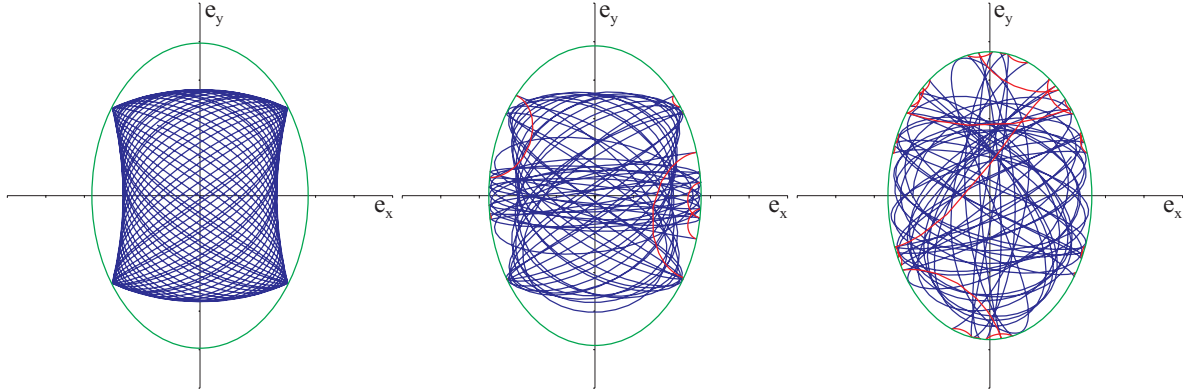}
\end{center}
\caption{The effect of relativistic precession on pyramid orbits
 \cite{MerrittVasiliev2010}.
The three orbits were  started with the same initial conditions,
but with different values of the coefficient $\kappa$
(equation~\ref{Equation:Definekappa}) that determines the relative speed of
relativistic and Newtonian precessions.
Left: $\kappa=0$ (regular);
middle: $\kappa=10^{-6}$ (weakly chaotic);
right: $\kappa=10^{-5}$ (strongly chaotic).
The outer ellipse marks the maximal extent of the ($u_x, u_y$) vector; 
red segments correspond to $\ell < \ell_\mathrm{crit}$,
blue to $\ell > \ell_\mathrm{crit}$ and to the nonrelativistic case.
} \label{Figure:exey}
\end{figure}

Addition of the 1PN relativistic terms to the equations of motion has a similar
effect on pyramids as on saucers: it limits the minimum angular momentum
attainable by a pyramid orbit, and defines a region very near the \sbh\ where
no pyramids can exist. 
Equations~(\ref{Equation:Definekappa}) -- (\ref{Equation:amincapture}), 
which were derived
for saucer orbits in axisymmetric nuclei, are approximately valid for pyramid
orbits if $\epsilon$ is identified with $\{\epsilon_b,\epsilon_c\}$.
One difference is integrability: pyramid orbits conserve only $E$ (rather than 
$E$ and $L_z$) in the presence of GR and so they tend to be chaotic, increasingly
so as $\kappa$ (equation \ref{Equation:Definekappa}) is increased, i.e. as the
distance from the \sbh\ decreases.
Instead of touching the equipotential surface at just four points,
the outer envelope of a chaotic ``pyramid'' orbit deforms to match the 
equipotential surface (Figure~\ref{Figure:exey}).
However one finds \cite{MerrittVasiliev2010} 
that the minimum attainable angular momentum is
not strongly affected  by the chaos and is still given by an equation similar
to  (\ref{Equation:ellminsaucer}).

In strongly triaxial nuclei, centrophilic orbits like the pyramids can dominate the orbital population of self-consistent models \cite{PoonMerritt2004}, and
the mass of stars on pyramid orbits 
can greatly exceed the mass on loss-cone orbits in the spherical or axisymmetric geometries. 
A reasonable estimate of the feeding rate in triaxial nuclei can be obtained by simply ignoring collisional loss-cone refilling and counting the rate at which stars on centrophilic orbits pass within a distance $r_\mathrm{lc}$ from the \sbh\ \cite{MerrittPoon2004}.
The neglect of relaxation is likely to be especially justified in the case of the largest
galaxies with the longest central relaxation times.

As in the case of saucer orbits in the axisymmetric geometry, 
pyramid orbits can precess past the  
loss cone in a time that is either
less than, or greater than, a radial period.
In the former case, the star has only a finite chance of capture
while in the latter case the star is guaranteed to pass through periapsis
before the orbit exits the loss cone.
The quantity that characterizes the two regimes is \cite{MerrittVasiliev2010}
\begin{equation}  \label{Equation:Define_qtri}
q_\mathrm{tri} \equiv \frac{\Delta\psi}{2\ell_\mathrm{lc}}
= \frac{P\nu_\mathrm{M}}{18\ell_\mathrm{lc}} \sqrt{W}
\end{equation}
where $\Delta\psi$ is the angle traversed in a radial period $P$.
If we consider a ``typical'' pyramid orbit  ($u_{x0}\approx u_{y0}$)
in a ``typical'' triaxial nucleus ($\epsilon_b\approx\epsilon_c$)
having peak angular momentum (near the $z$ axis) of $\ell_0$,
its draining time becomes \cite{MerrittVasiliev2010}
\begin{equation}
t_\mathrm{drain} \approx 
\cases {
\frac{1}{\sqrt{2}\;g(q_\mathrm{tri})}\frac{\ell_0}{\ell_\mathrm{lc}}t_\mathrm{pyr} & for 
$0\le q_\mathrm{tri} \le 1$, \cr
\frac{q_\mathrm{tri}}{\sqrt{2}} \frac{\ell_0}{\ell_\mathrm{lc}}t_\mathrm{pyr} & for $q_\mathrm{tri}>1$\cr}
\end{equation}
where 
\beq
g(x) \equiv \frac{2}{\pi}\left[\sqrt{1-x^2} + x^{-1}\sin^{-1}(x)\right] ; \ \ \ \ 
g(0)=\frac{4}{\pi}, \ \ \ \  g(1)=1 \nonumber
\eeq
and $t_\mathrm{pyr}$ is the period of a full libration cycle in $x$ or $y$,
which for eccentric pyramids is
\beq
t_\mathrm{pyr} \approx \frac{2\pi}{\nu_\mathrm{M}} \frac{1}{\nu_{x,y}}
\approx \frac{t_\mathrm{M}}{\sqrt{\epsilon}}.
\eeq
Since the maximal $\ell_0$ for pyramids is $\sim\sqrt{\epsilon}$,
the draining time in the ``empty-loss-cone'' regime ($q<1$) is a factor $\sim \sqrt{\epsilon}/\ell_\mathrm{lc}$ longer
than the pyramid precessional period, or $\sim \ell_\mathrm{lc}^{-1}$ longer than
the typical mass precession time $P\mh/M_\star$.
These inequalities reflect the fact that capture only occurs near the corners of the
pyramid, when oscillations in both $x$ and $y$ are simultaneously near their peaks,
and less often than once per full libration period in either $x$ or $y$.

The minimum angular momentum attainable in the presence of GR is expressible
in terms of $q_\mathrm{tri}$ by a relation similar to equation (\ref{Equation:ellminvsq}):
\beq
\frac{\ell_\mathrm{min}}{\ell_\mathrm{lc}} \approx \frac{3\pi}{\Theta} q_\mathrm{tri}^{-1}.
\eeq
Roughly speaking, the condition that stars be captured is equivalent to the statement that the loss cone is full. 

If $\eta_\mathrm{pyr}(E)$ is the fraction of stars at energy $E$ that are
on pyramid orbits, and $N(E) dE$ the total number of stars at energies $E$ to $E+dE$, 
the differential loss rate can be written approximately as
\begin{eqnarray}
\dot N(E) &\approx& \eta_\mathrm{pyr}N\; t_\mathrm{drain}^{-1}
 \nonumber \\
&\approx& \eta_\mathrm{pyr}\frac{M_\star}{\mh}\ell_\mathrm{lc}\frac{N}{P}
 \ \ \ \  \mbox{for }0\le q\le 1,  \nonumber \\
 &\approx& q_\mathrm{tri}^{-1}\eta_\mathrm{pyr}\frac{M_\star}{\mh}\ell_\mathrm{lc}\frac{N}{P}
\ \ \ \ \mbox{for }q>1. \label{Equation:Ndotpyramids}
\end{eqnarray}
(These expressions assume $\kappa=0$.)
Recall that in a spherical galaxy, the full-loss-cone capture rate
is $\sim \ell_\mathrm{lc}^2N/P$.
It is clear from equation~(\ref{Equation:Ndotpyramids}) that the loss rate
due to draining of the pyramids can be comparable to this.
Even though the time to drain one pyramid orbit is much longer than $P$,
the number of stars available to be captured in one draining time, $\eta_\mathrm{pyr}N$,
can be much larger than the number of stars on loss-cone orbits in a spherical galaxy, $\sim \ell_\mathrm{lc}^2N$.

After a time $\sim t_\mathrm{pyr}$
some parts of the orbital torus that are entering the loss regions
will be empty and the loss rate will drop below equation~(\ref{Equation:Ndotpyramids}).
For small $q_\mathrm{tri}$, the orbital torus will become striated,
containing strips of nearly zero density interlaced with
undepleted regions; the loss rate will exhibit discontinuous
jumps whenever a depleted region encounters a loss region
and the time to totally empty the torus will depend in a complicated
way on the frequency ratio $\nu_x/\nu_y$ and on $\ell_\mathrm{lc}$.
For large $q_\mathrm{tri}$, the loss rate will drop more smoothly with time,
roughly as an exponential law with time constant
$ t_\mathrm{drain}$.

In the case of pyramid orbits with arbitrary (not necessarily small) opening angles, 
numerical integrations suggest that the probability of finding an instantaneous
$\ell^2$ less than some value $X$ is given approximately by
$P(\ell^2 < X) \propto X$ for small $X$, corresponding
to a linear probability distribution of periapsis radii, 
$ P(r_\mathrm{peri}<r) \propto r$;
this is natural if one combines a quadratic distribution
of impact parameters at infinity with gravitational focusing
\cite{MerrittPoon2004}.
Defining $\mu$ for each orbit as
$P(\ell^2 < \ell_\mathrm{lc}^2)$, one finds that
while $\mu$ varies greatly from orbit to orbit, its overall distribution
over an ensemble of pyramid orbits is roughly
\begin{equation}  \label{mudistr}
P_\mu(\mu>Y) \approx \left(\frac{Y}{\mu_\mathrm{min}}\right)^{-2} \;, \qquad
\mu_\mathrm{min} \approx \frac{\ell_\mathrm{lc}^2}{2\tilde\eta}
\end{equation}
with $\tilde\eta$ the fraction of pyramids among all orbits.
The average $\mu$ for all pyramid orbits is therefore
$\overline{\mu} = 2\mu_\mathrm{min}$,
and the average fraction of time that an orbit of any $\ell$
spends inside the loss cone is
$\overline\mu \tilde\eta \simeq \ell_\mathrm{lc}^2$
(almost independent of the potential parameters $\epsilon_b$ and $\epsilon_c$)---the
same number that would result from an isotropic distribution of orbits
in a spherically symmetric potential.
In other words: until such a time as the centrophilic orbits have been substantially
depleted, loss-cone feeding rates should be roughly equal to full-loss-cone
rates in the equivalent spherical model.

\begin{figure}[!h]
\begin{center}
\includegraphics[width=0.80\textwidth,angle=0.]{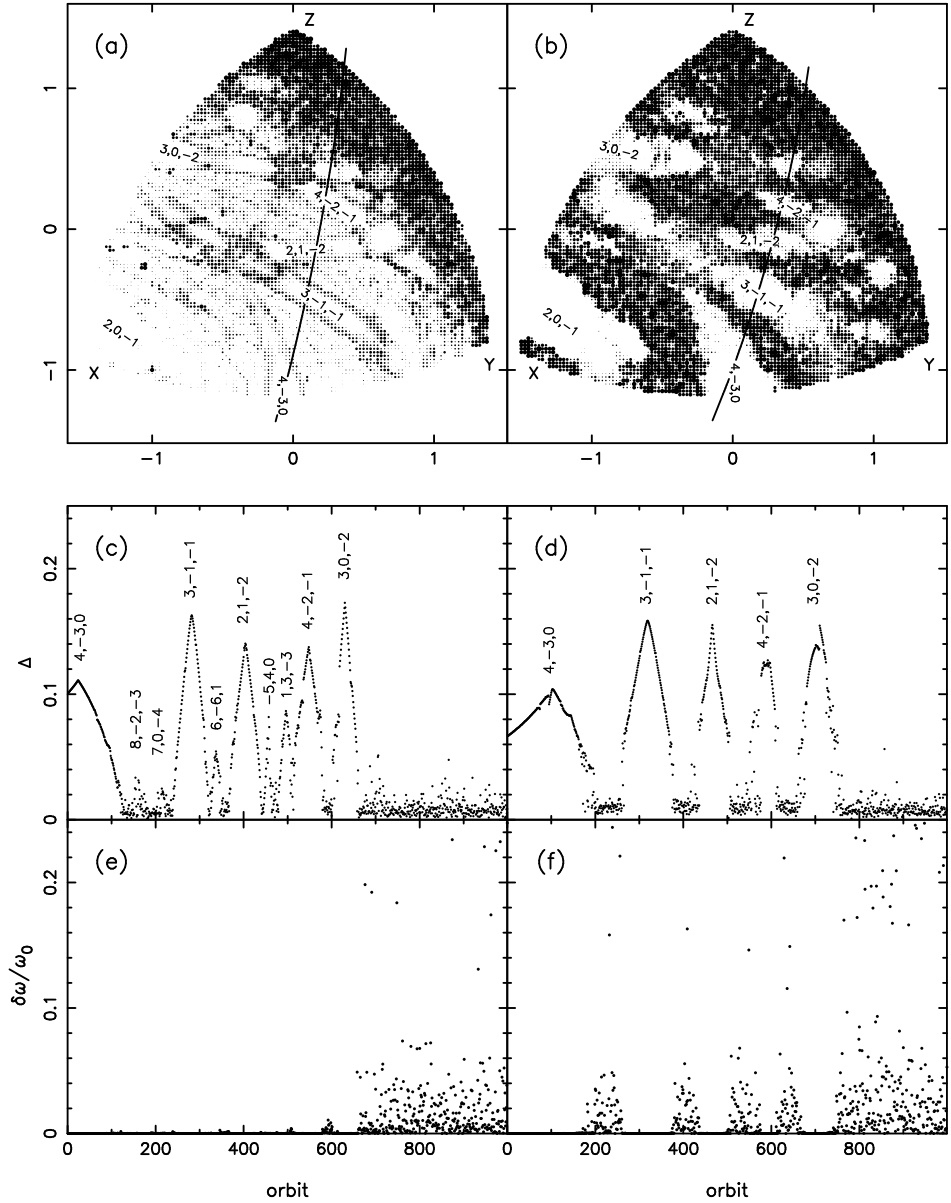}
\end{center}
\caption[]{Properties of centrophilic orbits in triaxial galaxies,
with (right) and without (left) central \sbhs\ \cite{MerrittValluri1999}.
The top panels show one octant of an equipotential surface located just
inside the half-mass radius of the model.
Orbits were started on this surface with zero velocity.
The top, left, and right corners correspond to the $z$- (short),
$x$- (long), and $y$- (intermediate) axes.
The gray scale is proportional to the logarithm of the
diffusion rate of orbits in frequency space;
initial conditions
corresponding to regular orbits are white.
The most important resonance zones are labeled with their
defining integers $(m_1,m_2,m_3)$.
Panels (c) and (d) show the distance of closest approach,
$\Delta$, of orbits whose starting
points lie along the heavy lines in (a) and (b).
The most important stable resonances are again labeled.
Panels (e) and (f) show the degree of stochasticity of the orbits,
as measured by the change $\delta\omega$ in their ``fundamental
frequencies'';
$\omega_0$ is the frequency of the long-axis orbit and regular orbits have $\delta\omega/\omega_0=0$.}
\label{Figure:rperitriax}
\end{figure}

As in the axisymmetric case, centrophilic orbits in the triaxial geometry
tend to become chaotic near the influence radius where radial and precessional
frequencies are comparable.
(As noted above, pyramid orbits are also strongly chaotic near 
the \sbh\ due to GR.)
In model potentials with a substantial degree of triaxiality, 
a ``zone of chaos'' extends from a few times $r_\mathrm{m}$ 
outward to a radius containing a mass in stars of  $\sim 10^2\mh$
\cite{ValluriMerritt1998}.
There are, broadly speaking, two types of centrophilic orbit in this region: 
regular orbits that
avoid passing through the very center,
and chaotic orbits.
The former orbits lie near to a ``thin'' orbit, that is, an orbit that respects a
resonance between the fundamental frequencies:
\begin{equation}
m_1\nu_1 + m_2\nu_2 + m_3\nu_3 = 0
\end{equation}
with the $m_i$ integers.
If the parent, resonant orbit avoids the center, orbits that lie close to the resonant
torus will do so as well, passing no closer to the center than some minimum distance
$\Delta$ (figure~\ref{Figure:rperitriax}).
As the initial conditions move farther from the resonant torus, the orbit broadens,
causing it to approach more closely to the destabilizing center.
At some critical $\Delta$---typically much larger than $r_\mathrm{lc}$---the orbit
becomes chaotic.
To a good approximation, {\it all} orbits that pass through the very center
and that extend outward into the ``zone of chaos'' are chaotic.

The complexity of the orbits in this region mandates a brute-force, numerical
treatment of \sbh\ feeding \cite{MerrittPoon2004}.
Such calculations are model-dependent but they suggest feeding rates
in triaxial galaxies of order
\beq
\dot M \approx  10^{-5}\eta
\left({\rh\over 100\,\mathrm{pc}}\right)^{-5/2}
\left({\mh\over 10^8\,\msun}\right)^{5/2}\,\msun\,\mathrm{yr}^{-1}
\label{Equation:Feed1}
\eeq
with a weak dependence on the degree of triaxiality; here $\eta$ is the
fraction of orbits that are centrophilic.
Equation~(\ref{Equation:Feed1}) is based on a nuclear model in
which $\rho\sim r^{-1}$, not too different from what is observed
near the centers of bright elliptical galaxies.


\section{Massive perturbers}
\label{Section:EvolveMPs}

It is straightforward to show that the characteristic time for scattering of a test
star by a set of field stars having a range of masses can be written as
\begin{equation}\label{Equation:DefineTrrAgain}
t_r = {0.34\sigma^3\over G^2 \tilde m\rho\ln\Lambda},\ \ \ 
\tilde m \equiv \frac{\int n(m) m^2 dm}{\int n(m) m\, dm} = \rho^{-1}\int n(m) m^2 dm,
\eeq
where $n(m) dm$ is the number of field stars with masses in the interval $dm$ centered on $m$.
Equation~(\ref{Equation:DefineTrrAgain}) assumes that field stars of all masses 
have the same velocity distribution (with dispersion $\sigma$),
a defensible assumption if the system is less than one relaxation time old.
To a first approximation, the tidal event rates derived in the preceding sections
can be generalized to a range of stellar masses by changing the  mass $m_\star$
that appears in equation~(\ref{Equation:DefineTr}) to $\tilde m$ and computing
the loss rate independently for stars in each $\{m_\star, R_\star\}$ group.

In a stellar population with a ``normal'' initial mass function (IMF) and in which the most massive stars have evolved off the main sequence, $\tilde m\approx \msun/2$.
If a nucleus is young enough that most of its stars are still
on the main sequence, the value of $\tilde m$ can be much larger.
For instance, a Salpeter \cite{Salpeter1955} IMF with  $10^{-2}\msun \le m \le 10^2\,\msun$ yields $\tilde m\approx 10^1\,\msun$, corresponding to
a relaxation time that is roughly ten times shorter than in an evolved cluster
with the same mass density.

Consider as an extreme case a mass function consisting of stars and some set of 
``massive perturbers'' with individual masses much greater than 
those of stars.
We can write
\beq
n(m) = \left[n(m)\right]_\mathrm{star} + \left[n(m)\right]_\mathrm{MP}
\eeq
and
\numparts
\begin{eqnarray}
\rho\, \tilde m &=& \left[\int n(m)m^2 dm\right]_\mathrm{star} +
\left[\int n(m)m^2 dm\right]_\mathrm{MP} \\
&=& \rho_\mathrm{star}\tilde m_\mathrm{star} +
\rho_\mathrm{MP}\tilde m_\mathrm{MP}.
\end{eqnarray}
\endnumparts
The condition that the scattering time be determined by the  massive objects is
\beq
\left(\rho\tilde m\right)_\mathrm{MP} \gg \left(\rho\tilde m\right)_\mathrm{star} .
\eeq

Near the center of a galaxy, massive perturbers can include gas clouds with
masses up to and including those of giant molecular clouds (GMCs), and
star clusters, both open and globular.
While mean number densities are very small---roughly $10^{-5}$\,pc$^{-3}$ in the
case of GMCs---these objects are so much more massive than stars that they
can easily dominate the gravitational scattering inside any region large enough
to contain them.
The dominant contribution to $\rho\tilde m$ turns out to come from the GMCs;
in the region $1.5\,\mathrm{pc} \lap r\lap 5\,\mathrm{pc}$,
one finds \cite{Perets2007}
\beq\label{Equation:muGMC}
\left(\rho\tilde m\right)_\mathrm{GMC} \approx (20\textrm{--}2000) \left(\rho\tilde m\right)_\mathrm{star}.
\eeq
In order to estimate the effective reduction in the timescale for gravitational
scattering, we need also to take into account the large physical size of GMCs,
which implies a lower effectiveness of close encounters.
The Coulomb logarithm can be written as
$\ln \Lambda \approx \frac12\ln(1 +  p_\mathrm{max}^2/p_0^2)$ .
Numerical experiments \cite{Spinnato2003} suggest 
that $p_\mathrm{max}$ is roughly $1/4$ times the linear extent of the test star's orbit.
Setting this size to 10\,pc, and replacing $p_0$ by  $R\approx 5$\,pc
($1/2$ the physical size of a GMC),
one finds that $\ln\Lambda$ near the center of the Milky Way
decreases from $\sim 15$ in the case of star--star scattering to
$\sim 0.5$ in the case of scattering by GMCs.
Combined with equation~(\ref{Equation:muGMC}), this result suggests
that the effective timescale for gravitational scattering
near the Galactic center might be reduced by a factor
of $\sim 10^0$ to $\sim 10^2$ due to the presence of GMCs.

What would be the consequences of such a reduction?
The effect on the distribution of stars inside the influence radius
of \SgrA, $r<\rh \approx 2$--$3$\,pc, is likely to
be small: at these radii, velocity perturbations are due mostly to objects
within $\sim \rh$, a region that is not likely to contain a single
massive perturber.
But the supply of stars to the \sbh\ is dominated by
gravitational encounters that take place at larger radii.
In the absence of massive perturbers, the transition from empty-
to full-loss-cone regimes takes place roughly at $r\approx \rh$
in a nuclear star cluster like that of the Milky Way (figure~\ref{Figure:CK}).
Massive perturbers cannot increase the flux in the full-loss-cone regime,
but they could convert an empty loss cone into a full loss cone,
implying an increased rate of capture by the \sbh.

Equation~(\ref{Equation:Fflccases}) gives for the energy-dependent
loss-cone flux is given in the two regimes as
\begin{equation}\label{Equation:FflccasesAgain}\nonumber
F \approx F^\mathrm{flc}\times
\cases {
 q|\ln{\cal R}_\mathrm{lc}|^{-1}, & if $q\ll -\ln{\cal R}_\mathrm{lc}$, \cr
1, & if $q\gg -\ln{\cal R}_\mathrm{lc}$ \cr},
\end{equation}
where $F^\mathrm{flc}$ is the full-loss-cone flux,
${\cal R}_\mathrm{lc}\approx r_\mathrm{lc}/r$, and
$q \approx  P/(T_r{\cal R}_\mathrm{lc})$
with $P$ the orbital period.
A decrease in the effective value of $t_r$ due to massive perturbers would imply a larger
$q$, hence a smaller radius of transition to the full-loss-cone regime.
But the effect on the net rate of stellar captures is likely to be modest,
at least in a galaxy like the Milky Way, since the transition
radius even in the absence of massive perturbers is $r_\mathrm{crit}\approx\rh$,
and since massive perturbers will not significantly decrease the effective value of
$t_r$ inside $\rh$.

These arguments are modified somewhat in the case of the interaction of {\it binary}
stars with an \sbh.
The distance from an \sbh\ at which a binary star is tidally separated,
$r_\mathrm{t,bin}$,
is larger than the tidal disruption radius for a single star, $r_\mathrm{t}$, by a factor
\beq
\frac{r_\mathrm{t,bin}}{r_\mathrm{t}} \approx \frac{a_\mathrm{bin}}{R_\star}
\approx 21 \left(\frac{a_\mathrm{bin}}{0.1\,\mathrm{AU}}\right)
\left(\frac{R_\star}{R_\odot}\right)^{-1}
\eeq
where $a_\mathrm{bin}$ is the binary semimajor axis.
Identifying $r_\mathrm{t,bin}$ with $r_\mathrm{lc}$ allows us to define a ``capture sphere''
for binary stars; the rate of diffusion of binaries into this sphere determines the rate
at which, for instance, hypervelocity stars are produced \cite{Hills1988}.
Since the capture sphere for binaries is so much larger than $r_\mathrm{t}$,
the empty-loss-cone regime extends much farther out, and any mechanism
that decreases the effective relaxation time can therefore have a substantial
effect on the binary disruption rate.
It has been argued that massive perturbers increase
the rate of interaction of binary stars with the
Milky Way \sbh\ by a factor of $10^1$--$10^3$,
with corresponding increases in the rate of production of hypervelocity stars,
and the rate of deposition of stars in tightly bound orbits near the \sbh\
\cite{Perets2007}.

\section{Relativistic loss cones}
So far only the lowest-order (1PN) relativistic corrections to the equations of motion
have been considered. Higher-order PN terms, representing the effects of frame dragging, of torques due to the \sbh's quadrupole moment, etc. can also be included
in the orbital equations \cite{MAMW2010}; of course, these higher-order corrections become progressively more important at smaller distances from the \sbh. 
But sufficiently close to the \sbh, the number of stars enclosed within any orbit is so small that it may no longer make sense to represent the gravitational potential from the stars as a smooth, symmetric function of position. Instead, the nonsphericity of the potential may be due mostly to the fact that at any moment, there are different numbers of stars on one side of the \sbh\ as compared with another. 
The idea is that---at least for some span of time---orbits near the \sbh\ are nearly Keplerian, and maintain their orbital elements, including particularly their orientations. The magnitude of the torque acting on a test star in this regime is roughly
\beq\label{Equation:RRTorque}
|{\cal T}| \approx \sqrt{N}\; \frac{Gm_\star}{a},
\eeq
where $N$ is the number of (field) stars inside the test-star's orbit, 
whose semimajor axis is $a$, and $m_\star$ is the mass of one field star.
Furthermore, since orbital periods $P$ are generally much smaller than
the time for orbital elements to change, it is reasonable to compute the torques
by averaging each orbit with respect to mean anomaly; in effect, replacing
each star by an elliptical ring of mass.\footnote{A similar averaging technique
is the basis for the derivation of the orbit-averaged Fokker-Planck equation  (section \ref{Section:SphericalSteadyState}); the Kozai-Lidov oscillations; 
and the results presented in sections 
\ref{Section:NonsphericalAxisym}---\ref{Section:NonsphericalTriax}
 for motion in axisymmetric and triaxial potentials, among many other 
examples in the literature.}

This ``$\sqrt{N}$ torque'' will dominate the torque from the large-scale nuclear nonsphericity if
\beq
N(a) \lap \epsilon^{-2}
\eeq
with $\epsilon$ defined in equation (\ref{Equation:Defineepsilon}).
So, for instance, if the nucleus is modestly elongated, $\epsilon \sim 10^{-1}$,
then at radii where $N(<a)\lap 10^2$, the $\sqrt{N}$ torques will dominate torques
from the large-scale distortion.
In the Milky Way, the corresponding radius might be $\sim 10^{-3}-10^{-2}$ pc
depending on the poorly-known distribution of stars and stellar remnants at these radii \cite{HopmanAlexander2006L,Merritt2010}.
One can obtain an approximate understanding of the motion in this regime by supposing
that the $\sqrt{N}$ torques are representable approximately in terms of an
axisymmetric or triaxial distortion (say), with amplitude given by $\epsilon \approx 1/\sqrt{N}$, and applying the results derived above.
What makes the problem much more interesting, 
and difficult, is the fact that the
orbits generating the torque do not maintain their orientations forever: they precess,
causing the direction of the torque generated by them
to change with time in some complicated way.

Evolution of orbits in response to $\sqrt{N}$ torques is called 
``resonant relaxation'' (RR) \cite{RauchTremaine1996}.
On time scales short compared with typical precession times, such
evolution is ``coherent'': the angular momentum of a test star increases 
approximately linearly with time in response to the nearly-constant torques.
On longer time scales, the direction of the $\sqrt{N}$ torques is changing, and
the response of a single orbit to the torques will be more like a random walk;
this is the ``incoherent'' regime.
The torques also affect the precession {\it rates}, but only slightly, and to a good
approximation, the rate at which stars precess can be computed by assuming
that the stellar potential is spherical.

But at these small distances from the \sbh, apsidal precession due to relativity 
typically can not be ignored,
and it affects the motion in two important ways.\footnote{Other effects of
relativity, including spin-orbit torques
and gravitational-wave energy loss, are considered below.}
(i) {\it Individual} stars on eccentric orbits can come close enough to the \sbh\ that
they experience relativistic precession in a time short compared with the time
for a typical star of the same $a$ to precess. 
This is similar to the behavior of a star on a saucer or pyramid orbit as it nears
the point of maximum eccentricity allowed by GR, as discussed above, and one expects that
the rate at which such a star random-walks in angular momentum space
(due to ``incoherent RR'') will drop precipitously when GR precession is so rapid
as to ``quench'' the effects of the $\sqrt{N}$ torques \cite{MerrittVasiliev2010}.
(ii) Sufficiently close to the \sbh, {\it most} stars will precess in response to GR
at a faster rate than precession due to other sources  (e.g. mass precession).
When this condition is met, GR sets the ``coherence time,'' the time
over which the $\sqrt{N}$ torques are nearly constant \cite{RauchTremaine1996}.

Taking the second of these first, we can define the coherence time,
$t_\mathrm{coh}$, at radius $a$ as the time for an orbit of typical
eccentricity to precess by an angle $\sim\pi$.
If precession is dominated by GR then equation (\ref{Equation:Sitter}) gives
\beq\label{Equation:tcohGR}
t_\mathrm{coh,S} (a) \approx \frac{1}{12} \frac{a}{\rg} P(a)
\eeq
while if mass precession is dominant,
\beq\label{Equation:tcohM}
t_\mathrm{coh,M}(a) \approx \frac{\mh}{N(a)m_\star} P(a).
\eeq
The coherence time is set by the shorter of these; 
$t_\mathrm{coh,S}$ is shorter than $t_\mathrm{coh,M}$ when
\beq
a N(a) \lap 12 \frac{\mh}{m_\star} \rg.
\eeq
In the Milky Way, the radius separating the two regimes is probably
$\sim 10^{-2}-10^{-1}$ pc.

As long as the ``test'' star is not precessing much faster than the ``field'' stars,
it feels a nearly constant torque over $\Delta t \lap t_\mathrm{coh}$, causing
its angular momentum to change by 
\beq\label{Equation:DefineTcohRR}
|\Delta L|_\mathrm{coh} \approx |{\cal T}| \Delta t \approx 
\sqrt{N}\;  \frac{Gm_\star}{a} \Delta t, \ \ \ \ \Delta t \lap t_\mathrm{coh}.
\eeq
On time scales longer than $t_\mathrm{coh}$, the $\sqrt{N}$ torques
from the field stars are changing direction in some complicated way, and
$|\Delta L|_\mathrm{coh}$ sets the step-length for a random
walk:
\beq\label{Equation:DLSRR}
|\Delta \boldsymbol{L}|\approx |\Delta \boldsymbol{L}|_\mathrm{coh,s}
\left(\frac{\Delta t}{t_\mathrm{coh}}\right)^{1/2}, \ \ \ \ 
\Delta t \gap t_\mathrm{coh}.
\eeq
We can write this as
\begin{equation}
\frac{|\Delta \boldsymbol{L}|}{L_c}  = \left(\frac{\Delta t}{t_\mathrm{RR}}\right)^{1/2},
\ \ \ \ 
t_\mathrm{RR} \equiv \left(\frac{L_c}{|\Delta\boldsymbol{L}|_\mathrm{coh}}\right)^2
t_\mathrm{coh}
\end{equation}
with $t_\mathrm{RR}$ the (incoherent) ``resonant-relaxation time'':
\beq\label{Equation:TRRmcoh}
t_\mathrm{RR}  \approx 
\cases {
\left(\frac{\mh}{m}\right)P, & $t_\mathrm{coh} = t_\mathrm{coh,M}$ \cr
\frac{3}{\pi^2} \frac{\rg}{a} \left(\frac{\mh}{m}\right)^2
\frac{P}{N}, & $t_\mathrm{coh} = t_\mathrm{coh,S}$ .\cr}
\eeq
These times can be compared to the time associated with non-resonant relaxation (NRR),
equation (\ref{Equation:DefineTr}), which can be rewritten approximately as\footnote{Note that $t_\mathrm{NRR}$ is the same quantity as $t_r$ defined above.}
\begin{equation}\label{Equation:trrapprox}
t_\mathrm{NRR} \approx \frac{C_\mathrm{NRR}}{\ln\Lambda} \left(\frac{\mh}{m}\right)^2\frac{P}{N} , \ \ \ \ C_\mathrm{NRR}\approx 0.1 \;.
\end{equation}
The condition $t_\mathrm{RR} < t_\mathrm{NRR}$ becomes
\numparts
\begin{eqnarray}
m_\star N(r<a) &\lap& \frac{\mh}{\ln\Lambda}, \ \ \ \ 
t_\mathrm{coh} = t_\mathrm{coh,M} \\
N(r<a) &\lap& C_\mathrm{NRR}^{1/2}\left(\frac{a}{\rg}\right)^{1/2}, \ \ \ \ 
t_\mathrm{coh} = t_\mathrm{coh,S} .
\end{eqnarray}
\endnumparts
At the Galactic center, the critical radius at which $t_\mathrm{RR}\approx t_\mathrm{NRR}$
is $\sim 0.05$ pc if there is a steeply-rising
density of stars and stellar remnants near the \sbh\ \cite{HopmanAlexander2006L}; 
or $\sim 0.2$ pc if the mass distribution is assumed to follow what is observed in
the bright, late-type stars, i.e. a central ``core'' \cite{Merritt2010}.
These radii are small compared with the radius at which most normal stars would
be scattered into the \sbh, whether by gravitational encounters, or by torques from
the large-scale mass distribution, and for this reason, resonant relaxation is typically
assumed not to strongly affect the event rates.

But the situation can be very different in the case of compact remnants. 
These include stellar-mass black holes (BHs) and neutron stars,
the end-states of the evolution of stars more massive than about $8\msun$.
The characteristic distance from which compact remnants would be scattered into 
a \sbh\ is smaller than for ordinary stars, for two reasons: remnants can
survive tidal disruption to much smaller separations; and---in the case of
stellar-mass BHs, which have masses $\sim 5\msun-20 \msun$---
mass segregation can cause them to accumulate near the \sbh.
The latter mechanism  occurs on a time scale of $\sim (m_\star/m_\bullet)t_\mathrm{NRR}$, where $m_\bullet$ is the BH mass.
This time may be less than $10$ Gyr at the Galactic center, and in ``relaxed''
models of the stellar distribution, the number (mass) density of
BHs  exceeds that of main-sequence stars at a distance of $\sim 10^{-2}$
($\sim 10^{-3}$) pc from the Milky Way \sbh\ \cite{HopmanAlexander2006L}.
However these models do not correctly reproduce the observed distribution
of normal stars (i.e. red giants), which exhibit a low-density core near the \sbh\
\cite{Buchholz2009,Do2009,Bartko2010}, 
suggesting that the relaxation time in the Galactic center
may not be short enough for a steady-state distribution to have been reached
\cite{Merritt2010}.

But suppose that the density in {\it some} nucleus (perhaps a nucleus containing
a \sbh\ less massive than the Milky Way's) is dominated by stellar remants
at radii $\ll \rh$, and that the resonant relaxation time  (the time for changing
orbital angular momenta) is much shorter, at these radii, than the non-resonant relaxation
time (the time for changing orbital energies).
Orbits of the remnants will undergo a random walk in $\boldsymbol{L}$ with characteristic
time $t_\mathrm{RR}$; since orbital energies remain nearly constant, changes in
$L$ correspond to changes in eccentricity $e$ and hence to changes in the radius of periapsis, $r_\mathrm{peri}=a(1-e)$.
In order to be captured by the \sbh, one of two things must happen.
(i) $r_\mathrm{peri}$ falls below
a few $\rg$;  e.g. for a non-spinning hole capture occurs if $r_\mathrm{peri} \lap 8\rg$ \cite{Will2012}.
Since $a\gg \rg$, the eccentricity of such a  capture orbit would be extremely high, 
$10^{-5}\lap 1-e \lap 10^{-3}$,
much higher than the average value $\langle e\rangle = 2/3$ for a ``thermal'' distribution
of orbits around a point mass.
(ii) The time scale for energy loss due to gravitational wave (GW) emission falls below
the time for orbital angular momenta to change, e.g., $(1-e) t_\mathrm{RR}$.
GW energy loss causes the semimajor axis of the orbit to shrink, initially at roughly constant
$r_\mathrm{peri}$ \cite{Peters1964}, until the orbit becomes nearly circular and 
the BH spirals into the \sbh.
The former channel is called a ``plunge'' and the latter an ``EMRI,'' or
extreme-mass-ratio inspiral \cite{AmaroSeoane2007,SigurdssonRees1997}.

The model as just described predicts interestingly high rates of capture 
\cite{HopmanAlexander2006},  but there is a problem.
As noted above, individual orbits that are highly eccentric will precess due to GR
at a rate much higher than the rates of precession of other stars with the same $a$.
In fact an eccentric orbit can precess so rapidly that the net effect of the $\sqrt{N}$ torques,
over one GR precessional cycle, is very small.
An equivalent way to say this is that---for a very eccentric orbit---the effective time over which the background torques can act coherently is given by {\it its} precession time, and not by  the average (and much longer) precession time, $t_\mathrm{coh}$, of the other orbits. 

The residual torque produced by an otherwise-spherical distribution
of stars is given by equation (\ref{Equation:RRTorque}).
Writing $L=\left[G\mh a(1-e^2)\right]^{1/2}$ for the angular
momentum of a test orbit,
the time scale over which the (fixed) torque changes $L$ is
\beq
\left|\frac{1}{L}\frac{dL}{dt}\right|^{-1} \approx
\sqrt{N(a)}\frac{\mh}{M_\star(a)}
\left[\frac{a^3(1-e^2)}{G\mh}\right]^{1/2}.
\eeq
(Note that we are comparing changes in $L$ to its own value, and not
to $L_c$.)
The condition that this time be shorter than the relativistic
precession time is\footnote{Due to a typesetter's error, the expression given for Eq. \ref{Equation:SB} in \cite{DEGN}, equation 6.195, is missing the square root on the 
left hand side.}
\beq\label{Equation:SB}
\sqrt{1-e^2} \gap \frac{\rg}{a} \frac{\mh}{m_\star}\frac{1}{\sqrt{N(a)}}\; .
\eeq
The same functional relation between $a$ and $e$ can be derived, 
up to a factor of order unity,  from expressions like
 (\ref{Equation:ellminsaucer}), the minimum angular momentum
attainable by a saucer or pyramid orbit in a fixed nonspherical potential,
after replacing $\epsilon$ by $1/\sqrt{N}$ \cite{MerrittVasiliev2010}.

\begin{figure}
\centering
\begin{tabular}{cc}
\includegraphics[width=1.\linewidth]{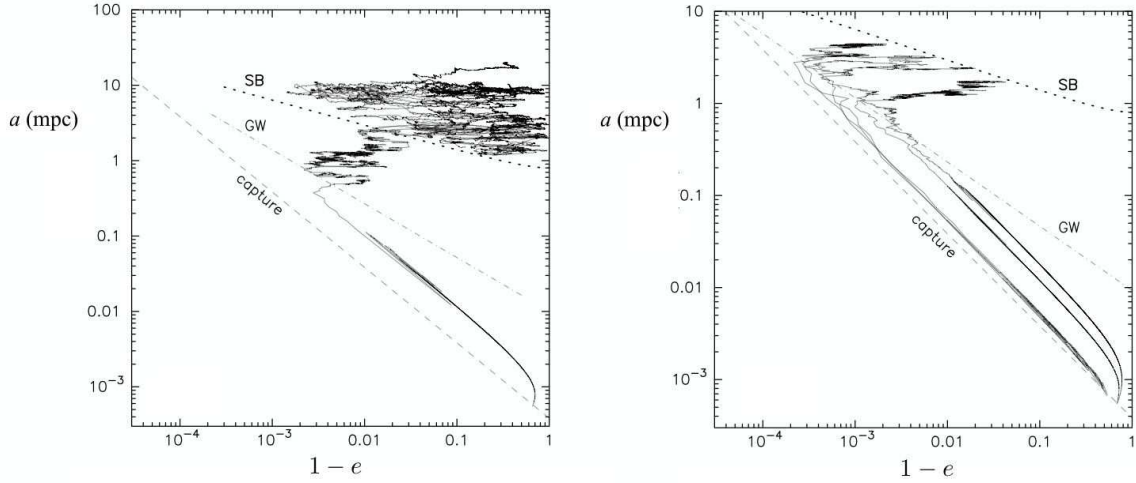}
\end{tabular}
\caption{Relativistic $N$-body simulation of EMRI formation \cite{MAMW2011}.
The left panel shows the trajectories, over a time interval of 2\,Myr,
of stellar-mass black holes orbiting a $10^6\,\msun$ \sbh\
as they undergo gravitational encounters with each other.
Motion in the $(a,e)$ plane is mostly horizontal  due to the fact that resonant
relaxation causes changes in angular momentum (i.e., $e$) on a timescale
that is much shorter than non-resonant relaxation causes changes in energy (i.e., $a$).
The dashed line marked ``capture'' is the capture radius around the \sbh; the
dotted line marked ``SB'' is  equation (\ref{Equation:SB});
and the dot-dashed line marked ``GW'' indicates the locus in the $a$--$e$ plane where angular momentum loss due to gravitational radiation dominates changes due to gravitational encounters.
Only one object in the left panel manages to cross the GW line and  become an EMRI;
most of the other objects are reflected by the eccentricity barrier
before reaching the gravitational-wave regime.
There are no ``plunges.''
The right-hand panel, a montage from several independent $N$-body integrations,
 shows a number of EMRI events like the single event in the left-hand panel.
}
\label{Figure:EMRI}
\end{figure}

The curve (\ref{Equation:SB}) is plotted as the dotted lines on figure~\ref{Figure:EMRI},
which is taken from the $N$-body study in which the phenomenon was discovered
\cite{MAMW2011}.
The $N$-body results confirm that stars are reluctant to cross this ``Schwarzschild
barrier'' (SB), which defines a locus of maximum eccentricity in the $(a,e)$ plane.
The barrier is predicted to exist for orbits with semimajor axes between $a_\mathrm{min}$ and $a_\mathrm{max}$:
the first value is obtained by setting $e=0$ in equation (\ref{Equation:SB}),
the second by the intersection of that relation with the capture
line, $r_\mathrm{peri} = a(1-e) = \Theta\rg\approx 8\rg$.
These limits can be expressed approximately as
\begin{eqnarray}\label{Equation:Defineamax}
\left(\frac{a_\mathrm{min}}{\mathrm{mpc}}\right)^2
\left(\frac{N_\mathrm{min}}{10^2}\right) &\approx& 0.2 
\left(\frac{\mh}{10^6\msun}\right)^{4}
\left(\frac{m_\star}{10\msun}\right)^{-2} , \nonumber \\
\label{Equation:Defineamin}
\left(\frac{a_\mathrm{max}}{\mathrm{mpc}}\right)
\left(\frac{N_\mathrm{max}}{10^2}\right) &\approx&  300 
\left(\frac{\Theta}{8}\right)^{-1}
\left(\frac{\mh}{10^6\msun}\right)^{3}
\left(\frac{m_\star}{10\msun}\right)^{-2}
\end{eqnarray}
where mpc $\equiv 10^{-3}$ pc and $N_{\{\mathrm{min,max}\}}$ is the
number of stars (or BHs), of mass $m_\star$, inside radius $r=\{a_\mathrm{min}, a_\mathrm{max}\}$.
If we adopt equation (\ref{Equation:SIS}) for $\rho(r)$---the ``singular isothermal sphere''---these relations become
\begin{eqnarray}
a_\mathrm{min} &\approx& 0.35 \left(\frac{\mh}{10^6\msun}\right)^{4/3} 
\left(\frac{m_\star}{10\msun}\right)^{-1/3}
\left(\frac{\sigma}{100\;\mathrm{km s}^{-1}}\right)^{-2/3} \mathrm{mpc}\; , \nonumber \\
a_\mathrm{max} &\approx& 8.0 \left(\frac{\Theta}{8}\right)^{-1/2}
\left(\frac{\mh}{10^6\msun}\right)^{3/2} 
\left(\frac{m_\star}{10\msun}\right)^{-1/2}
\left(\frac{\sigma}{100\;\mathrm{km s}^{-1}}\right)^{-1} \mathrm{mpc} \nonumber \\
\label{Equation:Defineaminamax}
\end{eqnarray}
or
\begin{eqnarray}
a_\mathrm{min} &\approx& 0.44 
\left(\frac{\mh}{10^6\msun}\right)^{1.2} 
\left(\frac{m_\star}{10\msun}\right)^{-1/3}
\mathrm{mpc}\; , \nonumber \\
a_\mathrm{max} &\approx& 11 \left(\frac{\Theta}{8}\right)^{-1/2}
\left(\frac{\mh}{10^6\msun}\right)^{1.3} 
\left(\frac{m_\star}{10\msun}\right)^{-1/2}
\mathrm{mpc} \nonumber \\
\end{eqnarray}
if the $\ms$ relation \cite{FerrareseMerritt2000} is used to
eliminate $\sigma$.
These scalings suggest that the relative extent of the barrier, $a_\mathrm{max}/a_\mathrm{min}$, is nearly
independent of $\mh$ and $m_\star$:
\beq
\frac{a_\mathrm{max}}{a_\mathrm{min}} \approx 25 \left(\frac{\Theta}{8}\right)^{-1/2}
\left(\frac{\mh}{10^6\msun}\right)^{0.1} 
\left(\frac{m_\star}{10\msun}\right)^{-1/6} .
\eeq
Note that these relations would only be expected to hold in galaxies containing
nuclear star clusters; galaxies having $\mh\gap 10^8\msun$ tend to exhibit central
cores, not NSCs, as discussed above.
The estimated value of $a_\mathrm{max}$ is small compared with \sbh\ influence radii 
(equation \ref{Equation:Definerh}), and so the existence of the SB is not likely to have
much consequence for the rate of tidal disruption of normal stars.
But the barrier can be expected to play a critical role in mediating the capture
of stellar remnants, or in determining the steady-state distribution of any other
nuclear component that can resist tidal disruption (dark matter particles, 
planetesimals, etc.)  at radii $r\lap a_\mathrm{max}$ 
\cite{HooperLinden2011,Zubovas2012}.

An orbit that ``strikes the barrier'' from above (i.e. from a region of lower
eccentricity) might be expected to behave---for a time less than $\sim t_\mathrm{coh}$---in a manner similar to orbits in fixed, axisymmetric or triaxial potentials 
when their angular momenta reach values close to the minimum allowed
by GR precession (e.g. equation~\ref{Equation:ellminsaucer}; figure~\ref{Figure:Axisym}).
In other words: $L$ should oscillate near its minimum value, at roughly the
GR precessional frequency.
That is roughly what is observed to happen in the $N$-body experiments \cite{MAMW2011};
but after remaining near the barrier for a few GR precessional
periods, orbits are observed to ``bounce back'' to lower values of $e$, 
where RR is once again effective.
What makes the behavior of orbits in an $N$-body system more complex than
in a fixed potential is, of course, that the
potential  generating the $\sqrt{N}$ torques is not constant: the field
star orbits also precess, and after a time $\sim t_\mathrm{coh}$, the direction of the torques
(and to a lesser extent, their magnitude) will have changed in some complex, but essentially 
random, way.

Near and below the barrier, orbits precess with frequency 
$\sim |d\omega/dt|_\mathrm{S}$ (equation \ref{Equation:Sitter}).
Precession in the
nearly fixed $\sqrt{N}$ potential results in a periodic variation in the test
star's angular momentum $\ell\equiv L/L_c = \sqrt{1-e^2}$:
\begin{eqnarray}
\ell(t) \approx \langle\ell\rangle \left[1-C \times \langle\ell\rangle \cos(\nu t)\right] ,\ \ \ \
\nu = \frac{3\left(G\mh\right)^{3/2}}{c^2 a^{5/2}\langle\ell\rangle^2}
\end{eqnarray}
where $C=C(a)$ is a poorly-determined ``constant'' of order
\beq
C\approx \frac{\sqrt{N}}{2} \frac{m_\star}{\mh} \frac{a}{\rg} .
\eeq
The amplitude of these oscillations is $\ell_+ - \ell_- \approx
2C\langle\ell\rangle^2$; below the SB, the amplitude  drops rapidly,
as $\sim (1-e^2)$.
On time scales {\it longer} than $t_\mathrm{coh}$, one might expect that
changes in the direction and amplitude of the $\sqrt{N}$ torques would
add a random component to the otherwise periodic variations in $\ell$
\cite{MAMW2011}.
As of this writing, there does not exist a good theoretical description of how
orbits evolve in this regime: that is: in response to (time-dependent) $\sqrt{N}$ torques
below the SB.
However, Monte-Carlo simulations, based on a simple (perhaps too simple) 
Hamiltonian model  \cite{MAMW2011}, suggest that $\langle\ell\rangle$ would indeed undergo a random walk in this regime.
Since this evolution is not well described either as ``resonant relaxation'' (RR) or 
``non-resonant relaxation'' (NRR), a new name seems appropriate.
Here we will call it ``anomalous relaxation" (AR).

One reason it is difficult to study AR via $N$-body simulations is that
its effects tend to be obscured by those of NRR, and to a greater degree than
would be expected in real nuclei.
Even though $t_\mathrm{RR}\ll t_\mathrm{NRR}$ at these radii, it does not follow
that NRR can be ignored near or below the barrier, 
since the effects of the $\sqrt{N}$ torques are so strongly suppressed by the rapid GR precession.
In fact at any $a\in [a_\mathrm{min}, a_\mathrm{max}]$,
there will be some eccentricity above which changes in angular momentum 
due to NRR are larger than those due to the $\sqrt{N}$
torques.
The time to change $L$ by of order itself due to NRR is
\beq\label{Equation:t1}
t_1 \approx \ell^2 t_\mathrm{NRR}(a) 
\approx 0.2 (1-e)  \left(\frac{\mh}{m}\right)^2\frac{P}{N\ln\Lambda} 
\end{equation}
(equation \ref{Equation:trrapprox}).
Let $t_2$ be the time for changes in $L$ due to AR, that is, due to the 
$\sqrt{N}$ torques alone, in the regime below the SB where GR precession is rapid.
While the $L$-dependence of $t_2$ is currently unknown, one expects $t_2$
to be an increasing function of $e$, since the increasingly rapid GR precession
below the barrier implies an increasing degree of adiabatic invariance of the
orbit's elements, and this expectation is at least qualtitatively consistent with 
results from the crude Hamiltonian
model mentioned above.
One estimate of the ratio of the two times, along the SB, is \cite{MAMW2011}
\begin{eqnarray}\label{Equation:DefineQ}
\left|\frac{t_1}{t_2}\right|_\mathrm{SB} &\approx&
\frac{1}{4C^2} \frac{t_\mathrm{NRR}}{t_\mathrm{coh}} 
\approx \frac{1}{N(a)} \left(\frac{\mh}{m_\star}\right)^2 \left(\frac{\rg}{a}\right)^2
\frac{t_\mathrm{NRR}(a)}{t_\mathrm{coh}(a)} \nonumber \\
&\approx& \frac{0.1}{\ln\Lambda} \left[\frac{\mh}{M_\star(a)}\right]^3
N^2(a)\left(\frac{\rg}{a}\right)^2, \ \ \ \ a\in [a_\mathrm{min}, a_\mathrm{max}]
\end{eqnarray}
where the latter expression sets $t_\mathrm{coh} = t_\mathrm{coh,M}$.
This scaling (if correct) suggests that $t_1/t_2\propto N^2$
in a nuclear model of specified $\mh$ and mass density.
Since the $N$-body simulations \cite{MAMW2011,Brem2013} 
have so far adopted unrealistically small values
of $N$ (i.e. too-large values of $m_\star$), one expects that they have been
affected by NRR to a greater degree than in real nuclei,
and hence that the simulated rates of capture were too high.
Before reliable estimates of capture rates can be made, simulations with
substantially larger $N$ will need to be carried out.
This is an extremely demanding problem from a computational point of view,
and a significant increase in the efficiency of $N$-body algorithms will probably
be needed in order to achieve this.

\begin{figure}
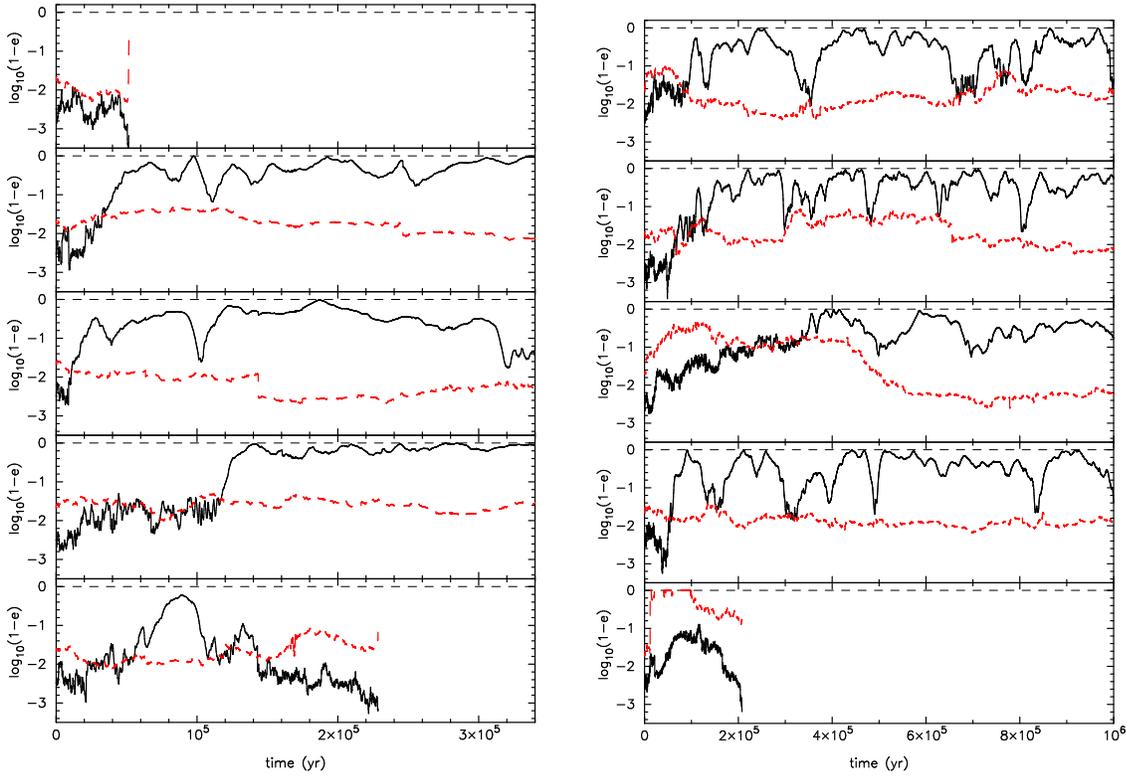

\begin{minipage}[b]{0.45\linewidth}
\centering
\includegraphics[width=\textwidth]{Merritt_Fig15a.eps}
\end{minipage}
\hspace{0.5cm}
\begin{minipage}[b]{0.45\linewidth}
\centering
\includegraphics[width=\textwidth]{Merritt_Fig15b.eps}
\end{minipage}
\caption{Eccentricity evolution of stars started ``below'' the 
Schwarzschild barrier (SB)  in a relativistic $N$-body integration,
including (left)  and without (right) the 2.5PN terms.
Solid (black) lines show $\log(1-e)$ and dotted (red) lines show the 
predicted eccentricity of the SB, equation~(\ref{Equation:SB}); the latter changes
with time due to changes in the stars semi-major axis $a$.
The lines terminate when the star is captured by the \sbh, 
which occurs in three of the panels.
Note two phenomena associated with the SB: the ``bounce'' that occurs when a star
strikes the barrier from above; and the ``buoyancy'' exhibited by a star that
crosses the barrier from below.\label{Figure:Buoy}}
\end{figure}

Another important question that will require large-$N$ simulations
is the steady-state distribution of stars or stellar remnants, 
with respect to angular momentum, near the SB. 
Imagine starting from a nucleus in which the region below the barrier is
unpopulated; ignore the effects of GW energy loss (appropriate if the
test bodies are of sufficiently low mass).
From time to time, penetration of the barrier from above will 
place a star below, where it will remain for a long time, since the time scale
for changes in $L$ below the barrier is long compared with the RR time
(figure \ref{Figure:Buoy}).
Eventually, the number of stars below the barrier will be so large that the
rate of passage from below to above will equal the rate from above to below.
The net rate of feeding of stars to the \sbh\ will be set by this steady-state
distribution.

While existing simulations probably do not have large enough $N$ to 
derive this distribution, there are other interesting questions whose 
answers do not depend on knowing the steady-state $N(L)$.
One such question is the origin of the S-stars, the bright, upper-main-sequence
stars that are observed to populate the inner $\sim 0.1$ pc of the Galactic center
\cite{Krabbe1995}.
These stars follow orbits with a nearly ``thermal'' distribution of eccentricities,
$N(e)\, de \sim e\, de$ \cite{Gillessen2009a};
whereas the leading model for their origin posits that they
were deposited on orbits of much higher eccentricity via exchange interactions
involving pre-existing binary stars \cite{Perets2007}.
Since the ages of the S-stars are less than about $100$ Myr, one needs to explain
how such extremely eccentric orbits evolved to their more modest eccentricities
in a relatively short time.
Purely Newtonian simulations \cite{Perets2009} suggested that RR could achieve
this, at least in nuclei with a sufficiently (and probably unphysically large) density
of compact-object perturbers.
But positing initially very eccentric orbits for the S-stars would almost certainly
place them below the SB, and their subsequent evolution would be something
like that of the test-particles shown in figure \ref{Figure:Buoy}.
After some time (which turns out to be of order $10^7$ in the Galactic center),
such a star may cross the SB from below to above, at which point RR ``turns on'' and carries
it to much higher angular momenta (lower $e$) in a time $\sim t_\mathrm{RR}$.
This ``buoyancy'' effect can clearly be seen in several panels of figure \ref{Figure:Buoy}
and it is a plausible explanation for the current $N(e)$ distribution of the S-stars \cite{AntoniniMerritt2013}.
While this is not strictly speaking a loss-cone problem, it does show that
the SB can be important for the evolution of the orbits of normal stars, even
stars that are directly observed.

Efforts are currently underway to build instruments capable of 
carrying out infrared astrometry to 10 micro-arc-second accuracy for stars near
the Milky Way SBH \cite{Eisenhauer2011}; one goal is to observe
deviations from Keplerian motion over time spans of a few years in the orbits
of stars with semimajor axes somewhat smaller than that of S2.
For these stars, the major sources of evolution are likely to be relativistic (Schwarzschild) precession of the periapsis, equation (\ref{Equation:Sitter}), and mass precession,
equation~(\ref{Equation:DefineMassPrecession}).
In addition, the $\sqrt{N}$ torques will induce changes in all the other
Kepler elements, at the rate defined above for ``coherent resonant relaxation,''
equation (\ref{Equation:DefineTcohRR}).
So, for instance, the changes over one orbital period of the eccentricity, $\Delta e$, 
and the direction of the orbital angular momentum vector, $\Delta\theta$, 
would be given by relations like
\numparts
\begin{eqnarray}\label{Equation:dedtS2a}
|\Delta e| &\approx& C_e\sqrt{N}\frac{m_\star}{M_\bullet},
\label{Equation:RRS2Ke} \\ \label{Equation:dedtS2b}
\Delta \theta &\approx& 2\pi C_t \sqrt{N}\frac{m_\star}{M_\bullet},
\label{Equation:RRS2Kt}
\end{eqnarray}
\endnumparts
where $N$ is understood to be the average number of stars inside the apoapsis of the orbit.
Numerical experiments \cite{Sabha2012} confirm these predictions and
allow the coefficients $\{C_e,C_t\}$ to be calibrated.
Because the changes in the star's orbit due to the $\sqrt{N}$ torques 
scale differently with $m_\star$ and $N$ than the changes due to the smoothly-distributed mass ($\propto M_\star=Nm_\star$), 
both the number and mass of the perturbing
objects within the observed star's orbit can in principle be independently constrained
\cite{Sabha2012}. 
For instance, one could determine $M_\star$ by comparing the observed
apsidal precession with the relativistic contribution (\ref{Equation:Sitter}),
then compute $m_\star\sqrt{N}$ by measuring changes in $e$ or $\theta$
and comparing with equations (\ref{Equation:dedtS2a})-(\ref{Equation:RRS2Kt}).

The relativistic effects described so far are a consequence of the lowest-order
(1PN) corrections to the Newtonian equations of motion.
What about the higher-order terms?
Energy loss due to GW emission is first reproduced at 2.5PN order, and in fact
the simulations shown in figures \ref{Figure:EMRI} and \ref{Figure:Buoy} 
 included the 2.5PN terms \cite{MAMW2011}.
If the \sbh\ is spinning (as it almost certainly is), there are additional 
corrections to the test-mass equations of motion  at low PN order.
These spin-orbit (Lense-Thirring, Kerr) terms imply an additional degree of apsidal precession of an orbiting star; 
this precession can usually be ignored compared with the Schwarzschild precession.
But the torques from a spinning hole also have a component which causes the
orbital line of nodes, $\Omega$, to precess, thereby changing the direction of the
orbital angular momentum vector $\boldsymbol{L}$.
The (orbit-averaged) rates of precession of a test mass's orbit due to the \sbh\
spin are given by
\numparts
\label{Equation:fOoSO}
\begin{eqnarray}
\left\langle\frac{d\Omega}{dt}\right\rangle_\mathrm{K} &=& \frac{2G^2\mh^2\chi}{c^3a^3(1-e^2)^{3/2}} = \frac{2G{\cal S}}{c^2a^3(1-e^2)^{3/2}}, \\
\left\langle\frac{d\omega}{dt}\right\rangle_\mathrm{K} &=& -\frac{6G^2\mh^2\chi}{c^3a^3(1-e^2)^{3/2}}\cos i
= -\frac{6G{\cal S}}{c^2a^3(1-e^2)^{3/2}} \cos i
\end{eqnarray}
\endnumparts
where $\chi\equiv cS/(G\mh)$ is the dimensionless spin, $i$ is the 
inclination of the star's orbit with respect to the \sbh's equatorial plane,
and $\Omega$ is defined also with respect to that plane.

\begin{figure}[!h]
\begin{center}
\begin{tabular}{cc}
\includegraphics[width=0.80\linewidth]{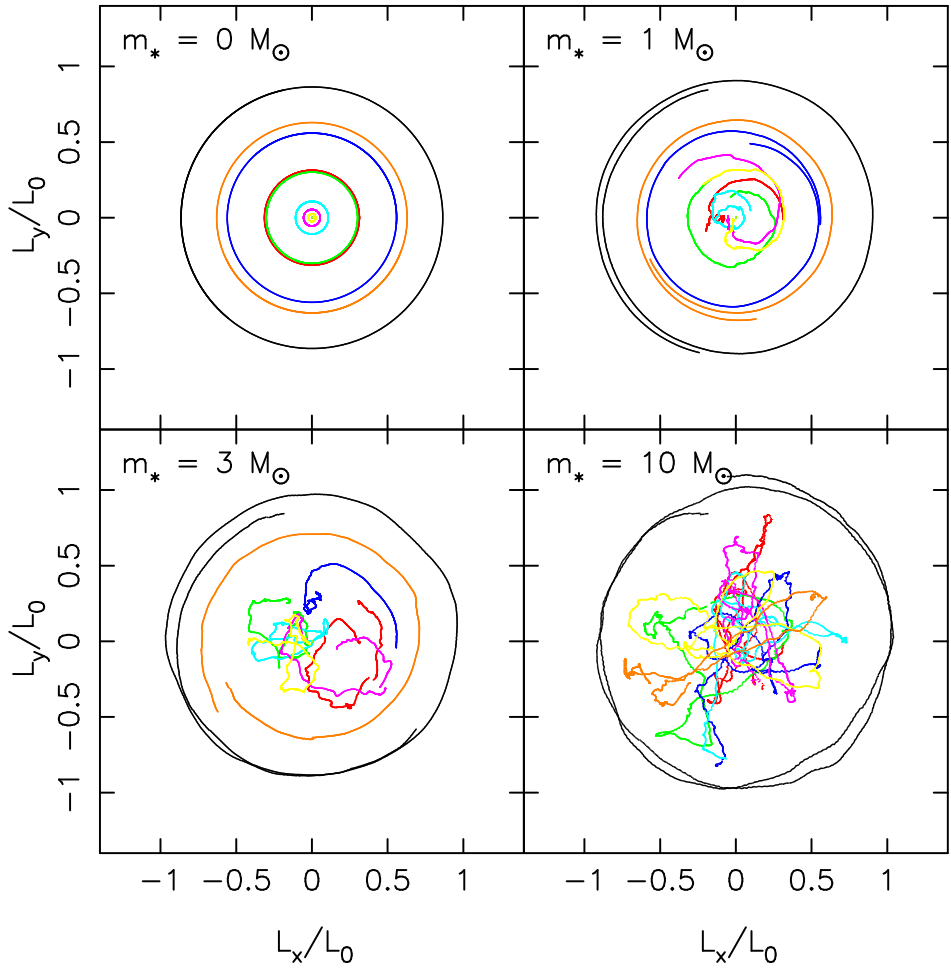}
\end{tabular}
\end{center}
\caption{
Evolution of orbital planes in an $N$-body simulation of a 
cluster of eight stars orbiting about the Galactic center \sbh\
($\mh=4\times 10^6\msun$) for an elapsed time of 
$2\times 10^6$ yr \cite{MAMW2010}.
The \sbh\ rotates about the $z$-axis with maximal spin.
Four different values were assumed for the stellar masses $m_\star$,
as indicated.
Stars were placed initially on orbits with semimajor axis
$a=2$\,mpc and eccentricity $0.5$ and with random orientations.
In a nucleus containing stars of a given mass, the transition between
motion like that in the first and last panels occurs at the ``rotational
influence radius'' (\ref{Equation:KerrCondition}).
\label{Figure:Kerr}}
\end{figure}

Just as there is a locus in the $(a,e)$ plane where Schwarzschild (apsidal) precession
inhibits the ability of the $\sqrt{N}$ torques to change the magnitude of $\boldsymbol{L}$ 
(equation \ref{Equation:SB}), so is there another curve along which Kerr (nodal) precession
inhibits the ability of the $\sqrt{N}$ torques to change the {\it direction} of
$\boldsymbol{L}$.
The latter is given roughly by
\beq\label{Equation:KerrCondition}
\left(1-e^2\right)^3\left(\frac{a}{\rg}\right)^3 \approx \frac{16\chi^2}{N(a)}
\left(\frac{\mh}{m_\star}\right)^2,
\eeq
which for a power-law ($\rho\propto r^{-\gamma}$) distribution of field stars
can be written
\numparts
\begin{eqnarray}\label{Equation:DefineaKa}
&&\left(1-e^2\right)^3\left(\frac{a}{a_\mathrm{K}}\right)^{6-\gamma} \approx 1, \\
&&a_\mathrm{K} = \rg\left( 8\chi^2 \frac{\mh}{m_\star}\right)^{1/(6-\gamma)}
 \left(\frac{r_\mathrm{m}}{\rg}\right)^{(3-\gamma)/(6-\gamma)}
\label{Equation:DefineaKb}
\end{eqnarray}
\endnumparts
where $a_\mathrm{K}$ is the ``rotational influence radius'' of the \sbh\ 
\cite{MerrittVasiliev2012}.
For reasonable nuclear models, this radius is,
very roughly, $a_\mathrm{K}\approx 10^4\rg$: small compared with the
radii $\{a_\mathrm{min}, a_\mathrm{max}\}$ that define the SB, but large
compared with the capture radius around the \sbh.
At $r\lap a_\mathrm{K}$, orbits evolve ``collisionlessly'' in response to
the Lense-Thirring torques, while for $r\gap a_\mathrm{K}$, star-star interactions
cause orbital planes to undergo a random walk on the (coherent) RR time scale
(figure \ref{Figure:Kerr}).
Note that there is no ``barrier'' associated with the rotational torques because
at these small radii the $\sqrt{N}$ torques are essentially unable to change
orbital eccentricities due to the rapid apsidal precession.
Neverthless, to the extent that the capture condition for a spinning \sbh\ is
dependent on orbital orientations, 
the transition from orderly precession at $r\lap a_\mathrm{K}$ to
a random walk of the orbital elements at $a\gap a_\mathrm{K}$ undoubtedly
has consequences for the loss-cone problem, which however have yet to be worked out.

\ack
This work was supported by the National Science Foundation under grant no. AST 1211602 
and by the National Aeronautics and Space Administration under grant no. NNX13AG92G.
Much of the work described here, including some unpublished work, was carried out in collaboration with F. Antonini, A. Hamers,
S. Mikkola, S. Portegies Zwart, E. Vasiliev, C. Will, and particularly with
T. Alexander. I thank E. V. for supplying figure 7.
I thank the referees, H. Perets and D. Pfenniger, for comments which improved
the presentation.

\section*{References}
\bibliographystyle{plain}	
\bibliography{/Users/merritt/Biblio/Main}

\end{document}